\documentclass[letterpaper,titlepage,11pt]{article}

\pdfoutput=1

\usepackage{amssymb,amsmath,amsfonts}
\usepackage{epsfig}
\usepackage{ccaption}
\usepackage{graphicx}

\usepackage[
      colorlinks=true,
      linkcolor=blue,
      urlcolor=blue,
      filecolor=black,
      citecolor=red,
      pdfstartview=FitV,
      pdftitle={},
        pdfauthor={Roberto Emparan, Ryotaku Suzuki, Kentaro Tanabe},
        pdfsubject={},
        pdfkeywords={},
        pdfpagemode=None,
        bookmarksopen=true
      ]{hyperref}

\def\clock{{\count0=\time
           \divide\count0 60
           \ifnum\count0<10 0\fi\the\count0
           \multiply\count0 -60 \advance\count0 \time
           :\ifnum\count0<10 0\fi \the\count0
         }}
\newcommand{\timestamp}{{\small\vbox{\hbox{\tt\jobname.tex}
\hbox{\the\day/\the\month/\the\year, \clock}}}}

\newcommand{\ie}{{\it i.e.,\,}}
\newcommand{\eg}{{\it e.g.,\,}}

\newcommand{\lp}{\left(}
\newcommand{\rp}{\right)}

\newcommand{\mc}[1]{\mathcal{#1}}

\newcommand{\beq}{\begin{equation}}
\newcommand{\eeq}{\end{equation}}
\newcommand{\bea}{\begin{eqnarray}}
\newcommand{\eea}{\end{eqnarray}}
\newcommand{\beqa}{\begin{eqnarray}}
\newcommand{\eeqa}{\end{eqnarray}}

\newcommand{\sR}{\mathsf{R}}
\newcommand{\sr}{\mathsf{r}}
\newcommand{\sN}{\mathsf{N}}
\newcommand{\sg}{\mathsf{g}}
\newcommand{\sV}{\mathsf{V}}
\newcommand{\sD}{\mathsf{D}}
\newcommand{\cR}{\mathcal{R}}
\newcommand{\cP}{\mathcal{P}}
\newcommand{\cK}{\mathcal{K}}

\newcommand{\Or}{\mathcal{O}}

\newcommand{\ord}[1]{{\mathcal O}(#1)}

\numberwithin{equation}{section}

\begin{document}

\begin{titlepage}
\rightline{KEK-TH-1810, AP-GR-122} 
\rightline{OCU-PHYS-422, KUNS 2557, YITP-15-33}
\vskip 1.cm
\centerline{\LARGE \bf Effective theory of Black Holes} 
\medskip
\centerline{\LARGE \bf in the $1/D$ expansion} 
\vskip 1.cm
\centerline{\bf Roberto Emparan$^{a,b}$, Tetsuya Shiromizu$^{c,d}$, Ryotaku Suzuki$^{e}$,}
\centerline{\bf Kentaro Tanabe$^{f}$, Takahiro Tanaka$^{g,h}$}
\vskip 0.5cm
\centerline{\sl $^{a}$Instituci\'o Catalana de Recerca i Estudis
Avan\c cats (ICREA)}
\centerline{\sl Passeig Llu\'{\i}s Companys 23, E-08010 Barcelona, Spain}
\smallskip
\centerline{\sl $^{b}$Departament de F{\'\i}sica Fonamental, Institut de
Ci\`encies del Cosmos,}
\centerline{\sl  Universitat de
Barcelona, Mart\'{\i} i Franqu\`es 1, E-08028 Barcelona, Spain}
\smallskip
\centerline{\sl $^{c}$Department of Mathematics, Nagoya University, Nagoya 464-8602, Japan}
\smallskip
\centerline{\sl $^{d}$Kobayashi-Maskawa Institute, Nagoya University, Nagoya 464-8602, Japan}
\smallskip
\centerline{\sl $^{e}$Department of Physics, Osaka City University, Osaka 558-8585, Japan}
\smallskip
\centerline{\sl $^{f}$Theory Center, Institute of Particles and Nuclear Studies, KEK,}
\centerline{\sl  Tsukuba, Ibaraki, 305-0801, Japan}
\smallskip
\centerline{\sl $^{g}$Department of Physics,
Kyoto University, Kyoto, 606-8502, Japan}
\smallskip
\centerline{\sl $^{h}$Yukawa Institute for Theoretical Physics,
Kyoto University, Kyoto, 606-8502, Japan}
\smallskip
\vskip 0.5cm
\centerline{\small\tt emparan@ub.edu,\, shiromizu@math.nagoya-u.ac.jp,\, ryotaku@sci.osaka-cu.ac.jp,\,}
\centerline{\small\tt  ktanabe@post.kek.jp,\, tanaka@yukawa.kyoto-u.ac.jp}

\vskip 1.cm
\centerline{\bf Abstract} \vskip 0.2cm 
\noindent 
The gravitational field of a black hole is strongly localized near its horizon when the number of dimensions $D$ is very large. In this limit, we can effectively replace the black hole with a surface in a background geometry (\eg\ Minkowski or Anti-deSitter space). The Einstein equations determine the effective equations that this `black hole surface' (or membrane) must satisfy. We obtain them up to next-to-leading order in $1/D$ for static black holes of the Einstein-(A)dS theory. To leading order, and also to next order in Minkowski backgrounds, the equations of the effective theory are the same as soap-film equations, possibly up to a redshift factor. In particular, the Schwarzschild black hole is recovered as a spherical soap bubble. Less trivially, we find solutions for `black droplets', \ie\ black holes localized at the boundary of AdS, and for non-uniform black strings.

\end{titlepage}
\pagestyle{empty}
\small
\tableofcontents
\normalsize
\newpage
\pagestyle{plain}
\setcounter{page}{1}

\section{Introduction}

Recently it has been demonstrated that black hole physics can be efficiently solved in an expansion around the limit of large number of dimensions, $D\to\infty$ \cite{Asnin:2007rw,Emparan:2013moa,Emparan:2013xia,Emparan:2014cia,Emparan:2014jca,Emparan:2014aba,Emparan:2015rva}. In this limit there appears a near-horizon region that is universal for all non-extremal neutral black holes, and which encompasses the small radial extent $\sim r_0/D$ of the gravitational field outside of a horizon of radius $\sim r_0$ \cite{Emparan:2013xia}. One important feature is that the quasinormal spectrum of the black hole splits into modes of high frequency, $\omega\sim D/r_0$, and of low frequency $\omega\sim 1/r_0$ \cite{Emparan:2014cia,Emparan:2014aba,Emparan:2015rva,Dias:2014eua}. The latter are particularly interesting, since they are fully localized in the near-horizon region (where they are normalizable excitations) and are decoupled from the asymptotic `far zone' to all perturbative orders in $1/D$.
This split into two scales makes it natural to try to integrate out the high-frequency, short-distance degrees of freedom to obtain a fully non-linear effective theory of the long-wavelength, decoupled dynamics. Such a theory should capture the physics of black holes on lengths and timescales $\gg r_0/D$, hence allowing for fluctuations on scales comparable to the horizon radius, $\sim r_0$, or (as appropriate for black branes \cite{Asnin:2007rw,Emparan:2013moa}) $\sim r_0/\sqrt{D}$.

We can also motivate this effective theory in a more geometric fashion. In the limit $D\to\infty$  the gravitational field of the black hole vanishes outside the near-horizon region, and thus there is a neat separation between the black hole and the background where it resides (\eg Minkowski or (Anti-)deSitter spacetime). The black hole can then be effectively identified with a particular surface $\Sigma_B$ in this background. Since ultimately the properties of the black hole are dictated by Einstein's equations, it must be possible to derive from them a set of equations that $\Sigma_B$ must satisfy. These constitute the effective theory of the black hole in the large $D$ limit. 

The black hole is then described by a set of `collective coordinates' which specify the embedding of $\Sigma_B$ in the background spacetime, and which vary over scales much larger than $r_0/D$, where $r_0$ is a characteristic length of $\Sigma_B$. The approach to obtain the effective theory employs the parametric separation between the large radial gradients and the smaller temporal and spatial gradients along the horizon, which allows to solve the radial dependence of the Einstein equations. Then, the vector-constraints in the radial direction yield the effective equations for the embedding functions.
Readers familiar with other effective theories of black holes, in particular the fluid/gravity correspondence \cite{Bhattacharyya:2008jc} and the blackfold approach \cite{Emparan:2009at}, will recognize similarities here. They are all based on a parametric separation between the fluctuations that are transverse to the horizon and those that are parallel to it. However, since $D$ is a parameter of the theory instead of a parameter specific to some solutions, in principle the $1/D$ expansion allows to tackle a larger set of problems ---at the expense of possibly losing accuracy at finite values of $D$ or missing phenomena which are non-perturbative in $1/D$.

This effective theory is an important step in the program to understand gravity in the large $D$ limit. General Relativity in vacuum (possibly with a cosmological constant, and without compactified dimensions) is essentially a theory of black holes and gravitational waves. The large $D$ effective theory is a reformulation of the black hole sector of General Relativity in terms of membrane-like variables. The decoupling property of the near-horizon zone implies that, as long as its gradients remain much smaller than $D/r_0$, the effective membrane can not radiate any gravitational waves to the far zone at any perturbative order in the expansion. Conversely, no gravitational waves from the far zone of frequency $\ll D/r_0$  can interact with the effective membrane; and, while far-zone waves of frequency $\sim D/r_0$ or larger can penetrate to the near zone, they are perfectly absorbed by the horizon on a short timescale $\sim r_0/D$ and do not interact with the low-frequency modes of the effective membrane \cite{Emparan:2013moa}. So the two sectors of the theory ---black holes and gravitational waves--- appear to be well separated, with the coupling between them being non-perturbative in $1/D$. However, there do exist black holes that when $D\to\infty$ have large spatial gradients $\sim D/r_0$ along their horizons. Moreover, in the evolution of some horizons it can occur that initially small temporal gradients grow to values $\sim D/r_0$. Such situations imply breakdowns of the applicability of the effective theory, and are reminiscent of the breakdown of hydrodynamics when turbulence develops.

In this article we start to develop the large $D$ effective theory of black holes by focusing on the simplest case of static configurations of neutral black holes, possibly with a cosmological constant. The equations for the embedding of $\Sigma_B$ in the background take a remarkably simple form: if the trace of the extrinsic curvature of $\Sigma_B$ is $K$, and the redshift factor on $\Sigma_B$ is $g_{tt}$, then the effective equation at leading order is
\beq\label{effeq0}
\sqrt{-g_{tt}}\,K=2\kappa\,,
\eeq
where the constant $\kappa$ gives the surface gravity of the black hole.
When this is satisfied, our results provide explicitly the near-horizon black hole metric that solves the full Einstein-(A)dS equations in the leading large $D$ limit. Observe that in backgrounds like Minkowski space where there is no redshift, this is the same as the equation for a soap film. The stress-energy tensor of the effective membrane that lives on $\Sigma_B$ is also simple: it is a modulation along the membrane directions of the quasilocal stress tensor of the large-$D$ black hole.

We also obtain the next-to-leading order corrections to the effective theory. Although in general the form of the equations becomes a little more complicated than \eqref{effeq0}, in the absence of redshifts we get the same equation $K=\mathrm{const.}$ at the next order in the expansion.

In order to test the usefulness of the theory, we have applied it to obtain several non-trivial solutions. Staticity greatly restricts the number of possible black holes. For instance, in all dimensions the unique asymptotically flat, static black hole of Einstein's vacuum equations is the Schwarzschild-Tangherlini solution \cite{Gibbons:2002av}, which is recovered from \eqref{effeq0} as a spherical soap bubble --- correspondingly, the spherical soap bubble is the unique closed surface of constant mean curvature in flat space in any dimension\footnote{This is widely believed among differential geometers to be the case, see \eg\cite{hsiang}. We thank Gary Gibbons for this reference.}. However, we will show that the effective theory easily yields other static solutions: black holes localized at the boundary of AdS (`black droplets') \cite{Hubeny:2009ru}, and non-uniform black strings in asymptotically flat space \cite{Gregory:1993vy,Gubser:2001ac,Wiseman:2002zc}. 
The construction of these solutions at specific finite values of $D$ required sophisticated numerical solution of systems of partial differential equations. In contrast, the large-$D$ equations for these problems are single ordinary differential equations that can be easily solved, when not in an analytical exact or approximate form, at least with a one-line command of $\texttt{NDSolve}$ in Mathematica. 

In the next section we introduce a formalism that is adequate for the resolution of the problem. Then in sec.~\ref{sec:LOtheory} we solve the Einstein equations at leading order in $1/D$ and derive the effective equation for the surface $\Sigma_B$. We also discuss simple examples, and obtain the stress-energy tensor for the effective membrane on $\Sigma_B$. In sec.~\ref{sec:droplets} the effective equation is solved to obtain black droplets in AdS. Sec.~\ref{sec:NLOeq} contains the derivation of the effective theory at next-to-leading (NLO) order. This is then applied in sec.~\ref{sec:nubs} to the construction of non-uniform black strings. Sec.~\ref{sec:outlook} finishes the paper with some brief remarks. The appendices contain technical details and elaborations on asides mentioned in the main text.

\section{Set up}

In our metric ansatz we separate a radial direction $\rho$, where gradients are of order $D$, from all other spacetime directions along which variations are smaller. To this effect, we use a decomposition in `synchronous gauge'
\beq
ds^2=N^2(\rho,x)\frac{d\rho^2}{(D-1)^2}+g_{\mu\nu}(\rho,x) dx^\mu dx^\nu\,,
\eeq
where $\mu,\nu$ run over all the $D-1$
directions orthogonal to $\rho$. 
The Einstein equations in vacuum, with a cosmological constant
\beq \Lambda=-\frac{(D-1)(D-2)}{2\ell^2}\,,\eeq
can be written in terms of the intrinsic, $R_{\mu\nu}$, and extrinsic, $K_{\mu\nu}$, curvature tensors of the $(D-1)$-dimensional constant-$\rho$ surfaces,
\beqa
&&
K^2-K^\mu{}_\nu K^\nu{}_\mu=R+{(D-1)(D-2)\over \ell^2},
\label{Hconst}
\\ &&
\nabla_\nu K^{\nu}_{\mu}-\nabla_\mu K=0
\label{Mconst}
\\ &&
{D-1\over N}\partial_\rho K^\mu{}_\nu+K K^\mu{}_\nu = R^\mu{}_\nu
    +\delta^\mu{}_\nu \frac{D-1}{\ell^2}
 -{1\over N}\nabla^\mu\nabla_{\nu} N,
\label{evolEq}
\\ &&
 K^\mu{}_\nu ={D-1\over 2N}g^{\mu\sigma}\partial_\rho g_{\sigma\nu}\,.
\label{extcurv}
\eeqa
Eqs.~\eqref{Hconst} and \eqref{Mconst} are respectively the scalar and vector constraints, while \eqref{evolEq} is the `dynamical evolution' equation in the radial direction. Knowing $K^\mu{}_\nu$, eq.~\eqref{extcurv} can be integrated to obtain the metric. 
It is convenient to consider separately the equation for $K$ obtained from eq.~\eqref{evolEq}, 
\begin{equation}\label{treveq}
{D-1\over N}\partial_\rho K+K^2=(D-1)^2\cK^2(\rho,x)
\,, 
\end{equation}
where we define $\cK(\rho,x)$ 
by
\begin{equation}
\cK^2(\rho,x)=\frac{1}{\ell^2}+\frac1{(D-1)^2}\lp R-\frac{1}{N}\nabla^2 N\rp\,. 
\label{hatr}
\eeq

We take the metric to be static. In the spatial directions orthogonal to $\rho$, we have non-trivial dependence on a number $p=\ord{D^0}$ of spatial directions $z^a$, which are orthogonal to a $n+1$-dimensional space
that we take to be a sphere $S^{n+1}$.\footnote{This can be  extended to other spaces with intrinsic curvature $\propto n^2$ at large $n$.} We introduce
\beq\label{ndp}
n=D-p-3\,,
\eeq
which can be used as the large expansion parameter instead of $D$. 
Our metric ansatz is then
\beq\label{nhansatz}
ds^2=N^2(\rho,z)\frac{d\rho^2}{(D-1)^2}-\sV^2(\rho,z)dt^2+
\sg_{ab}(\rho,z)dz^a dz^b+\sR^2(\rho,z)q_{ij}dx^i dx^j
\eeq
where $a=1,\dots,p$, and $q_{ij}$ is the metric on the unit $S^{n+1}$. 

The different metric functions will be assumed to scale with $n$ in specific ways. In order to get oriented, note that one solution that we intend to recover is the leading-order near-horizon geometry of the Schwarzschild black hole \cite{Emparan:2013xia}, which can be written in the ansatz \eqref{nhansatz} with $p=1$ as\footnote{Here $\rho$ is twice the one in \cite{Emparan:2013xia}.}
\beqa\label{schwnh}
ds^2
&=&r_0^2\frac{d\rho^2}{n^2}-r_0^2 \tanh^2(\rho/2)dt^2+r_0^2\lp 1+\frac4{n}\ln\cosh(\rho/2)\rp\lp dz^2+\sin^2 z\, d\Omega_{n+1}\rp\,.
\eeqa
Observe that in the sphere radius we are keeping terms of order $1/n$: due to the large dimensionality of the sphere, such terms enter (through traces) in the leading order equations and thus must be kept at this order. The takeaway here is that, while the $\rho$-dependence in $\sV$ appears at the leading order, instead in $\sg_{ab}$ and $\sR$ it is at $\ord{1/n}$.

We then assume that
\beq\label{nscaling}
K^t{}_t\,,\, K=\ord{n}\,,\qquad K^a{}_b\,,\, K^i{}_j=\ord{1}\,,\qquad R=\ord{n^2}\,.
\eeq
We also assume that $N$, $\sV$, $\sR$ are $\ord{1}$
and that
\beq\label{assgdR}
\sg_{ab},\,\partial_a \sR =\ord{1},\ \mathrm{or}\ \ord{1/n}\,.
\eeq
In these two cases the Ricci tensor of $\sg_{ab}$ is, respectively,
\beq\label{assRab}
{}^{(\sg)}\!R^a{}_b=\ord{1},\ \mathrm{or}\ \ord{n}\,.
\eeq
The first instance in \eqref{assgdR} will be exemplified in sec.~\ref{sec:droplets}, and the second one in sec.~\ref{sec:nubs}.

\section{Effective theory: leading order}\label{sec:LOtheory}

Our strategy is to first solve eqs.~\eqref{Hconst} and \eqref{evolEq} for the radial dependence of the extrinsic curvature with regularity at the black hole horizon, and then obtain the metric from \eqref{extcurv}. Radial integrations leave an undetermined function of $z^a$ which cannot be eliminated by gauge choices: this is the one collective degree of freedom of the static black hole. The vector constraint \eqref{Mconst} then yields a non-linear differential equation, on $z$ only, for this degree of freedom, which is the effective equation we seek.

\subsection{Solving the leading order equations}

According to our assumptions we can separate the leading-order, $\rho$-independent terms as
\beqa
N(\rho,z)&=&N_0(z)+\ord{1/n},\,\\
\sg_{ab}(\rho,z)&=&\gamma_{ab}(z)+\ord{1/n},\,\\
\sR(\rho,z)&=&\mc{R}(z)+\ord{1/n},\\
\cK(\rho,z)&=&\frac1{r_0(z)}+\ord{1/n}\,.\label{defr0}
\eeqa
In this case we can integrate \eqref{treveq} to find\footnote{Henceforth in this section we omit the symbol $\ord{1/n}$ to unburden the notation. In addition, the subsequent analysis is valid only when $r_0^2>0$.} 

\begin{equation} 
 K=\frac{n}{r_0(z)}\coth \lp\frac{N_0(z)}{r_0(z)}(\rho-\rho_0(z))\rp\,. 
\end{equation}
The divergence at $\rho=\rho_0(z)$ is the expected pole at the horizon, coming from $K^t{}_t$ (all other components of the extrinsic curvature must be regular there). The metric to leading order is invariant under $\rho\to \rho+f(z)$, which we use to set $\rho_0(z)=0$. Similarly, by rescaling $\rho$ by an $\ord{1}$ function of $z^a$ we can reach a gauge in which, to leading order, 
\begin{equation}\label{gaugeN}
 N_0(z)=r_0(z)\,.
\end{equation}
Then
\begin{equation} 
 K=\frac{n}{r_0(z)} \coth {\rho}. 
\label{Keq}
\end{equation}

Next, the equation for $K^t{}_t$ is 
\begin{equation}
{n\over r_0(z)}\partial_\rho K^t{}_t+K K^t{}_t=0, 
\end{equation}
since $R^t{}_t +n/\ell^2$ can be neglected as it is of lower order.
This equation is solved as 
\begin{equation}
 K^t{}_t=\frac{n}{r_0(z) \sinh \rho},
\label{Ktt}
\end{equation}
where the integration function of $z$ has been fixed again by requiring that $K^t{}_t$ have a pole at  $\rho=0$. 

Since $K^a{}_b$ and $K^i{}_j$ are $\ord{1}$, in order to solve for them we need $R^a{}_b$ and $R^i{}_j$. The components of the curvature in a constant-$\rho$ section
\beqa
g_{\mu\nu}dx^\mu dx^\nu
&= &-\sV^2(\rho,z)dt^2+\gamma_{ab}(z) dz^a dz^b+\mc{R}^2(z)q_{ij}dx^i dx^j
\eeqa
are readily obtained (see appendix~\ref{app:rhocurv}). The scalar curvature gets negligible contributions from the time direction, and is given at leading order by
\beq
R=n^2\frac1{\mc{R}^2}\lp 1-(\sD\mc{R})^2\rp\,,\label{R1}
\eeq
where we abbreviate
\beq
(\sD\mc{R})^2=\gamma^{ab}\partial_a\mc{R}\,\partial_b\mc{R}\,.
\eeq
Comparing \eqref{R1} to eqs.~\eqref{hatr} and \eqref{gaugeN} we find
\beq\label{eqN}
\frac1{r_0^2(z)}=\frac1{\ell^2}+\frac1{\mc{R}^2}\lp 1-(\sD\mc{R})^2\rp\,.
\eeq
It is easy to see that this equation is actually equivalent to the scalar constraint \eqref{Hconst} at leading order. 

Since the scalar curvature is dominated by the components along the sphere we have
\beq
R^i{}_j=\frac1{n}\delta^i{}_j\, R=n\,\delta^i{}_j \lp\frac1{r^2_0}-\frac1{\ell^2}\rp\,.
\eeq

Using the covariant derivative $\sD_a$ for the metric $\gamma_{ab}$, the curvature along the $z^a$ directions, obtained from \eqref{Rab}, is 
\beq
R^a{}_b=-n\frac{\sD^a \sD_b \mc{R}}{\mc{R}}+{}^{(\gamma)}\!R^a{}_b\,.
\eeq 
Here we take into account that in the second case in \eqref{assgdR}, \eqref{assRab} the intrinsic curvature of $\gamma_{ab}$ can contribute at the same order.

Now we have all the terms needed to integrate the equations from \eqref{evolEq},
\beqa
\frac{n}{r_0}\partial_\rho K^a{}_b+K K^a{}_b&=&R^a{}_b+\frac{n}{\ell^2}\delta^a{}_b\,,\\
\frac{n}{r_0}\partial_\rho K^i{}_j+K K^i{}_j&=&R^i{}_j+\frac{n}{\ell^2}\delta^i{}_j\,.
\eeqa
Imposing regularity at $\rho=0$ we find
\beqa
K^a{}_b&=& r_0 f^a{}_b\tanh(\rho/2)\,,\label{Kzz}\\
K^i{}_j&=&\delta^i{}_j\,\frac1{r_0}\tanh(\rho/2)\label{Kij}\,,
\eeqa
where we have defined the tensor
\beq\label{fab}
f_{ab}(z)=\frac{\gamma_{ab}(z)}{\ell^2}-\frac{\sD_a\sD_b\mc{R}}{\mc{R}}+\frac{{}^{(\gamma)}\!R_{ab}}{n}\,.
\eeq

It is now straightforward to integrate the extrinsic curvatures \eqref{Ktt}, \eqref{Kzz} and \eqref{Kij} to obtain the metric components. This gives
\beqa
\sV(\rho,z)&=&V_0(z)\tanh(\rho/2)\,,\label{gttsol}\\
\sg_{ab}(\rho,z)&=&\gamma_{ab}(z)+\frac{4}{n}r_0^2(z)\,f_{ab}(z)\,\ln\cosh(\rho/2)\,,\label{gabsol}\\
\sR(\rho,z)&=&\mc{R}(z)\lp 1+\frac{2}{n}\ln\cosh(\rho/2)\rp\,.\label{Rsol}
\eeqa
$V_0(z)$ is an integration function from the $\rho$-integration of $K^t{}_t$. For the $ab$ and $ij$ components the integration functions of $z$ have been absorbed in a $\ord{1/n}$ redefinition of $\gamma_{ab}(z)$ and $\mc{R}(z)$. However, $V_0(z)$ cannot be absorbed in that way.

Up to this point we have solved all the radial dependence of the metric and it only remains to impose the vector constraint \eqref{Mconst}, whose only non-trivial component is along $z^a$. To leading order it takes the form
\beq
\Gamma^t_{at}K^t{}_t +\Gamma^j_{ai}K^i{}_j+\lp \partial_b\ln\sqrt{-g}\rp K^b{}_a +\sD_a K=0\,.
\eeq
Plugging in our previous results and using the identity 
\beq\label{ident}
(\partial_b\ln\mc{R})\lp \frac{\delta^b{}_a}{\ell^2}-\frac{\sD^b\sD_a\mc{R}}{\mc{R}}\rp=\frac1{r_0^2}\lp \partial_a\ln\mc{R}-\partial_a\ln r_0\rp \,,
\eeq
derived from \eqref{eqN}, we find, 
\beqa\label{veceq}
\partial_a\ln V_0(y)-\partial_a\ln r_0(z)-\frac{2r_0^2}{n} (\partial_b\ln\mc{R})\,{}^{(\gamma)}\!R^b{}_a\sinh^2(\rho/2)=0\,.
\eeqa
Under either of the two cases in \eqref{assgdR}, \eqref{assRab}, the last term is subleading and can be neglected. Thus, consistently, the $\rho$-dependence cancels out of the leading-order equation, which requires that $V_0$ be proportional to $r_0$. 
Observe that this condition is equivalent to requiring that the surface gravity $\kappa$ be uniform on the horizon. Indeed, the precise relation is
\beq\label{effeq1}
V_0(z)=\frac{2\kappa}{n} r_0(z)\,.
\eeq
Defining a rescaled surface gravity
\beq
\tilde\kappa=\frac{\kappa}{n}\,,
\eeq 
our solution of the Einstein equations, valid in the near-horizon region to leading order in $1/n$, is
\beqa\label{effsoln}
ds^2&=&r_0^2(z)\lp -4\tilde\kappa^2 \tanh^2(\rho/2)dt^2+\frac{d\rho^2}{n^2}\rp+\lp\gamma_{ab}(z)+\frac{4}{n}r_0^2(z)f_{ab}(z)\ln\cosh(\rho/2)\rp dz^a dz^b\notag\\
&&+\mc{R}^2(z)\lp \cosh(\rho/2) \rp^\frac{4}{n}d\Omega_{n+1}\,.
\eeqa
Although $\tilde\kappa$ could be absorbed by rescaling the time coordinate, this may not be convenient, since the normalization of $t$ is fixed by matching to the far-zone.

The geometry \eqref{effsoln} can be interpreted as a modulation along $z^a$ of a near-horizon Schwarzschild black hole solution. The functions $r_0(z)$, $\gamma_{ab}(z)$ and $\mc{R}(z)$ vary slowly compared to radial gradients, and one of them, say $\mc{R}(z)$, can be regarded as the single collective degree of freedom for the black hole. Indeed, it is possible to give an alternative derivation of the effective theory following these ideas, as described in appendix \ref{app:another}.

\subsection{Effective equation}\label{sec:effLO}

The metric \eqref{effsoln} in the near-horizon geometry must be matched to the far-zone background in the common `overlap zone', which corresponds to $1\ll \rho \ll n$ in \eqref{effsoln}. The matching must be such that the metric induced on a constant-$\rho$ surface $\Sigma_B$ there, namely,
\beq\label{sigmamet}
ds^2\bigl|_{\Sigma_B}=-V_0^2(z)dt^2+\gamma_{ab}(z)\,dz^a dz^b+\mc{R}^2(z)d\Omega_{n+1}\,,
\eeq
and its extrinsic curvature, are the same when approached from either zone. From \eqref{gttsol} and \eqref{Keq} we see that 
\beq
K\bigl|_{\Sigma_B}= \frac{n}{r_0(z)}\,,\qquad \sqrt{-g_{tt}}\bigl|_{\Sigma_B}=V_0(z)\,.
\eeq 
Therefore eq.~\eqref{effeq1} can be written in the simple form
\beq\label{effeq}
\sqrt{-g_{tt}}\,K\bigl|_{\Sigma_B}=2\kappa
\eeq
with constant $\kappa$. 

That is, if in a given background spacetime we find a surface that satisfies \eqref{effeq}, then we can `resolve' this surface by replacing it and its interior with the static black hole with geometry \eqref{effsoln}, whose surface gravity is $\kappa$.
Let us remark that $\sqrt{-g_{tt}}\,K$ in the limit to the horizon is, by definition, equal to $2\kappa$. The vector constraint is independent of $\rho$, but it is non-trivial that $\sqrt{-g_{tt}}\,K$ takes the same value on the horizon and on $\Sigma_B$.

Using \eqref{eqN} we can write \eqref{effeq}  more explicitly. For metrics on $\Sigma_B$ of the form \eqref{sigmamet}, the square of  \eqref{effeq} is 
\beq\label{effeqex1}
\frac{V_0^2(z)}{\mc{R}^2(z)}\lp 1-(\sD\mc{R})^2+\frac{\mc{R}(z)^2}{\ell^2}\rp=4\tilde\kappa^2\,.
\eeq
When there is only one $z$ coordinate, $p=1$, this equation takes the form
\beq\label{effeqex}
\frac{V_0^2(z)}{\mc{R}^2(z)}\lp 1-\gamma^{zz}\mc{R}'(z)^2+\frac{\mc{R}(z)^2}{\ell^2}\rp=4\tilde\kappa^2\,.
\eeq
This is the version that we will employ in the examples in this paper.

Finally, observe that in our construction of the black hole solution, $K$ is positive when the radial normal points outwards of the horizon. This resolves any possible ambiguity about which side of $\Sigma_B$ in the background spacetime corresponds to the exterior of the black hole: if $K\bigl|_{\Sigma_B}>0$, the exterior of the black hole lies in the direction of the normal to $\Sigma_B$.

\subsection{Simple solutions}\label{sec:ssoln}

We verify that the effective theory correctly reproduces known exact solutions.

\paragraph{Schwarzschild-(A)dS black holes.}

In global AdS spacetime (extending to Minkowski and deSitter when $1/\ell^{2}\leq 0$),
\beq
ds^2=-\lp 1+\frac{r^2}{\ell^2}\rp dt^2+\frac{dr^2}{1+\frac{r^2}{\ell^2}}+r^2\lp d\theta^2+\sin^2\theta d\Omega_{n+1}\rp,
\eeq
take a surface $\Sigma$ at $r=\bar{r}(\theta)$, so that
\beq
V_0^2(\theta)=1+\frac{\bar{r}^2(\theta)}{\ell^2}\,,\qquad \gamma_{\theta\theta} (\theta)=\bar {r}(\theta)^2+\frac{(\bar{r}'(\theta))^2}{1+\frac{\bar{r}(\theta)^2}{\ell^2}}\,,\qquad \mc{R}(\theta)=\bar{r}(\theta)\sin\theta\,.
\eeq
It is immediate to see that a spherical surface
\beq
\bar{r}=r_h
\eeq
solves \eqref{effeqex} with the correct surface gravity
\beq
\tilde\kappa=\frac1{2r_h}\lp 1+\frac{r_h^2}{\ell^2}\rp.
\eeq 
The static planar and hyperbolic black holes in AdS are similarly easy to obtain. 

In a Minkowski background, $1/\ell=0$, we can also find the same solution starting from
\beq\label{minkback}
ds^2=-dt^2+dz^2+dr^2+r^2d\Omega_{n+1}\,.
\eeq
Setting $r=\bar{r}(z)$, so that $\gamma_{zz}=1+(\bar{r}')^2$ and $\mc{R}=\bar{r}$,  eq.~\eqref{effeqex} is solved by $\bar{r}=\sqrt{r_h^2-z^2}$ with $\bar\kappa=1/(2r_h)$. Thus the Schwarzschild solution when $D\to\infty$ is obtained as a spherical soap bubble with mean curvature
\beq
\frac{1}{n}K=\frac1{r_h}.
\eeq 

\paragraph{Black strings.} Both in the Minkowski background \eqref{minkback} and in Poincar\'e AdS,
\beq\label{PAdS}
ds^2=\frac{\ell^2}{z^2}\lp -dt^2+dr^2+r^2d\Omega_{n+1}+dz^2\rp,
\eeq 
the black string is just $r=r_h$, with $\tilde\kappa=1/(2r_h)$.

\paragraph{Intersecting black hole and deSitter horizons.}
Let us describe a less simple, lesser-known solution in the deSitter background,\footnote{We thank Jorge Santos for discussions that led to this example.} with $H^2=-\ell^{-2}>0$, 
\beq\label{desit}
ds^2=-\lp 1- H^2 r^2\rp dt^2+\frac{dr^2}{1- H^2 r^2}+r^2\lp d\theta^2+\sin^2\theta d\Omega_{n+1}\rp\,.
\eeq
We change coordinates $(r,\theta)\to (\tilde{r},\tilde{\theta})$ as
\beq
\sin\tilde{\theta}=H r\sin\theta\,,\qquad \tilde{r}\cos\tilde{\theta}=r\cos\theta\,,
\eeq
so that \eqref{desit} becomes
\beq\label{desit2}
ds^2=\cos^2\tilde{\theta}\lp -\lp 1- H^2 \tilde{r}^2\rp dt^2+\frac{d\tilde{r}^2}{1- H^2 \tilde{r}^2}+\tilde{r}^2d\Omega_{n}\rp+\frac1{H^{2}}\lp d\tilde{\theta}^2+\sin^2\tilde{\theta} d\phi^2\rp.
\eeq
In this case,
\beq\label{tildrh}
\tilde{r}=r_h
\eeq
is a solution of \eqref{effeqex} (where $z=\tilde{\theta}$ and $\cR(\tilde\theta)=r_h\cos\tilde{\theta}$), with
\beq
\tilde\kappa=\frac{1}{2r_h}\lp 1-H^2 r_h^2\rp\,.
\eeq
This black hole is actually known exactly in all $D$ \cite{Emparan:2011ve}: it is the metric
\beq\label{bhds}
ds^2=\cos^2\tilde{\theta} \lp -fdt^2+\frac{d\tilde{r}^2}{f}+\tilde{r}^2d\Omega_{n}\rp+\frac1{H^{2}}\lp d\tilde{\theta} ^2+\sin^2\tilde{\theta}  d\phi^2\rp
\eeq
with 
\beq
f=1-\lp\frac{r_0}{\tilde{r}}\rp^{n-1}- H^2 \tilde{r}^2\,,
\eeq
which arises in the Kerr-deSitter family of black holes in the limit in which the equator of the rotating black hole touches the deSitter horizon and the configuration becomes static. The two horizons correspond to the cosmological and black hole horizons of the submetric $(t,\tilde{r},\Omega_n)$, which extend separately along the coordinate $\tilde\theta$ until they meet at the equator $\tilde\theta=\pi/2$. It is a simple exact instance of a horizon intersection. By the same reasoning as used in \cite{Emparan:2013moa}, in the large $n$ limit the black hole of \eqref{bhds} becomes a `hole' at $\tilde{r}=r_h=r_0+\ord{1/n}$ in the geometry \eqref{desit2}.

In the coordinates of \eqref{desit}, the solution \eqref{tildrh} is the surface
\beq\label{staticMP}
r=\bar{r}(\theta)=\frac{r_h}{\sqrt{1-\sin^2\theta\lp 1- H^2 r_h^2\rp}}\,.
\eeq
If we write
\beq\label{xyr}
x=\bar{r}\cos\theta\,,\qquad y=\bar{r}\sin\theta\,,
\eeq 
then \eqref{staticMP} becomes the ellipse
\beq
\frac{x^2}{r_h^2}+ H^{2}y^2=1\,.
\eeq
Since $\bar{r}(0)=r_h<1/H$ and $\bar{r}(\pi/2)=1/H$, the surface $\bar{r}(\theta)$ is an oblate ellipsoid that touches the deSitter horizon at its equator, in agreement with our interpretation above.

\subsection{Effective stress tensor}\label{effstensor}

In order to match the near-zone solution to the far zone, they must share the metric \eqref{sigmamet} at their common boundary (in the overlap zone), and also their extrinsic curvatures. In this way the full geometry is smoothly glued between the two zones. This construction is equivalent to substituting the near-zone by a membrane with geometry \eqref{sigmamet} and with a stress-energy tensor given by the quasilocal stress-energy tensor
of the near-horizon geometry measured in the overlap zone. Having this membrane stress-energy tensor as a source for the gravitational field in the background, ensures that the matching of the near- and far-zone geometries is $C^1$.

The overlap zone is the asymptotic region of the near-zone.
Asymptotically at large $\rho$, and neglecting terms $\ord{e^{-\rho}}$, the metric \eqref{effsoln} becomes
\beqa\label{asymet}
ds^2&\to&-V_0^2(z)dt^2+r_0^2(z)\frac{d\rho^2}{n^2}+\lp\gamma_{ab}(z)+\frac{2\rho}{n}r_0^2(z)f_{ab}(z)\rp dz^a dz^b\notag\\
&&+\mc{R}^2(z)\lp 1+\frac{2\rho}{n}\rp d\Omega_{n+1}\,.
\eeqa
with the condition \eqref{effeq1}. When the scalar constraint \eqref{eqN} is satisfied, this is in fact a solution at all $\rho$: it corresponds to empty space (Minkowski or AdS) at large $n$. Therefore \eqref{eqN} is not specific to black holes. Instead, it pertains to the definition of near-zone asymptotics of large $n$ gravity.

The geometry \eqref{asymet} is then the reference metric required to define the effective stress tensor. In general there can be terms in the metric at large $\rho$ of the type $ a_{\mu\nu}(z)\,\rho/n$, but all the functions $a_{\mu\nu}(z)$ that appear are fully determined, up to gauge, by the Einstein equations in the overlap zone in terms of $r_0(z)$ and $\mc{R}(z)$ and their derivatives. In this sense, they are analogous to the first terms in the Fefferman-Graham (FG) expansion in AdS, which are fixed by the boundary metric. The terms at large $\rho$ that behave like $e^{-\rho}$, $e^{-2\rho}$\dots, correspond to normalizable perturbations in the near-zone and are not determined by the boundary metric. The leading terms $\propto e^{-\rho}$ will give the quasilocal stress-energy tensor.

To leading order at large $\rho$, the quasilocal stress-energy tensor from \eqref{asymet} is the same for empty space and for the black hole. Thus this stress tensor cannot give any non-trivial gravitational effect on the background. Such effects come from the difference between the quasilocal stress energy tensors of empty space and of the black hole, \ie\ 
\beq\label{effstress}
8\pi G\, T^\mu{}_\nu =-[K^\mu{}_\nu]+\delta^\mu{}_\nu [K]
\eeq
where the subtraction in $[K^\mu{}_\nu]=K^\mu{}_\nu -{}^{(0)}K^\mu{}_\nu$ is from empty space with the same asymptotic boundary metric. The subtraction removes the terms $\propto\rho$ at large $\rho$, and leaves those that decay like $e^{-\rho}$ (or faster). Typically, in the far zone we have $e^{-\rho}\sim (r_0/r)^n$, which is the fall-off of the gravitational field away from a localized source.

For the black hole solution \eqref{effsoln}, expanding at large $\rho$, neglecting terms $\Or(e^{-2\rho})$, and
subtracting the background values we find
\beqa
[K^t{}_t]&=&\frac{2n}{r_0(z)}e^{-\rho}+\Or(n^0)\,,\\{}
[K^a{}_b]&=&-2r_0(z)f^a{}_b\, e^{-\rho}+\Or(1/n)\,,\\{}
[K^i{}_j]&=&-\delta^i{}_j\frac{2}{r_0(z)} e^{-\rho}+\Or(1/n)\,,\\{}
[K]&=&\Or(n^0)\,.
\eeqa
It is easy to check that
\beq
\nabla_\mu[K^\mu{}_a]-\partial_a[K]=\ord{n^0}
\eeq
since this indeed follows from the vector constraint. It implies the conservation of the stress-energy tensor \eqref{effstress}, which is given by 
\beqa
8\pi G\, T^t{}_t&=&-\frac{2n}{r_0(z)}e^{-\rho}+\Or(n^0)\,,\notag\\
8\pi G\, T^a{}_b&=&\Or(n^0)\,,\\\
8\pi G\, T^i{}_j&=&\Or(n^0)\,,\notag
\eeqa
with the constraint that
\beq\label{Tconstr}
n (\partial_b\ln \cR)T^b{}_a-(\partial_a\ln \cR)\delta^j{}_i T^i{}_j= T^t{}_t (\partial_a \ln r_0)+\Or(n^0)\,.
\eeq
That is, at this order the values of $T^b{}_a$ and $T^i{}_j$ are not fully determined, since terms of $\ord{n^0}$ in $[K]$, which remain undetermined at this order, can make a contribution to these  components. However, the combination in \eqref{Tconstr} does not suffer from this indeterminacy and this equation is required for the conservation of the stress-energy tensor at leading order. It is not surprising that the tensions (or pressures) play a subleading role: it was already observed in \cite{Emparan:2013moa} that the pressure of black branes is suppressed relative to their energy density when $D\gg 1$.

This effective stress tensor is enough to obtain the backreaction of the black hole on the background, by computing the linearized gravitational field created by this source. The specific values of $T^a{}_b$ and $T^i{}_j$ are not important at this order, so one may choose them arbitrarily as long as \eqref{Tconstr} (which is necessary for a consistent coupling of the source to the gravitational field) is satisfied.
While these components may be determined at the next order in the expansion, one way of choosing them is by requiring that, as is the case in all known black hole solutions, the `hoop stresses' $T^i{}_j$ vanish. Under this assumption we have
\beqa
8\pi G\, T^b{}_a\,\partial_b\ln \cR(z)&=&-\frac{2}{r_0(z)}\partial_a\ln r_0(z) e^{-\rho}+\Or(1/n)\,,\label{otherT}\\
8\pi G\, T^i{}_j&=&0\,.
\eeqa
In the case $p=1$ we can solve \eqref{otherT} to find
\beq
8\pi G\, T^z{}_z=-\frac{2}{r_0(z)}\frac{\partial_z\ln r_0(z)}{\partial_z\ln \cR(z)\,} e^{-\rho}+\Or(1/n)\,.
\eeq

\subsubsection{Effective membrane source}\label{sec:effmem}

Since the stress tensor is homogeneous on $S^{n+1}$, we can integrate it over the angular directions. At large $\rho$ this yields a factor $\Omega_{n+1}\mc{R}^{n+1} e^{\rho}/4$. Then the integrated energy density is
\beq\label{intT}
\langle T ^t{}_t\rangle=-\frac{n\Omega_{n+1}}{16\pi G }\frac{\mc{R}(z)^{n+1}}{r_0(z)}+\Or(n^0)\,,
\eeq
and the stresses are, as mentioned above, $\ord{n^0}$ and therefore subleading but must satisfy
\beq
\langle T^b{}_a\rangle\,  \partial_b\ln \cR(z)=-\frac{\Omega_{n+1}}{16\pi G }\frac{\mc{R}(z)^{n+1}}{r_0(z)}\partial_a\ln r_0(z) +\Or(1/n)\,.
\eeq
When $p=1$ we can explicitly solve this to obtain 
\beq\label{intTzz}
\langle T^z{}_z\rangle=-\frac{\Omega_{n+1}}{16\pi G }\frac{\mc{R}(z)^{n+1}}{r_0(z)}\frac{\partial_z\ln r_0(z)}{\partial_z\ln \cR(z)}+\Or(1/n)\,.
\eeq

The energy density \eqref{intT} admits a very simple physical interpretation: it is the energy density of a black brane of radius $\mc{R}(z)$, redshifted by the local redshift factor $1/r_0(z)$. The tension \eqref{intTzz} is also the tension for that type of black brane, with an additional correction that accounts for the bending tension.

The backreaction of the black hole on the background is obtained by taking a membrane with this stress-energy tensor, extended along the directions $z^a$ in the background, and at the origin of the $S^n$: by homogeneity in the sphere direction, in linearized gravity and at large $n$, we need not consider that the membrane extends on a sphere of finite radius but rather we can collapse it to zero radius.

In appendix~\ref{app:farSchw} we describe how the field created by this stress tensor correctly yields the large-$n$ linearized Schwarzschild field, when it is interpreted as a solution of \eqref{effeqex} in the background \eqref{minkback}. 

\section{Black droplets}
\label{sec:droplets}

Now we construct solutions of \eqref{effeqex} in the AdS spacetime \eqref{PAdS} that are rather more complicated than in sec.~\ref{sec:ssoln}. We seek surfaces $\Sigma_B$ that end at the boundary at $z=0$ on a sphere $S^{n+1}$, and extend a finite distance into the bulk at $z>0$ until the $S^{n+1}$ shrinks to zero. 
They correspond to black holes localized at the AdS boundary. These were investigated in the context of cutoff-AdS/CFT (the holographic duality for the Randall-Sundrum-II braneworld) for the purpose of studying Hawking radiation of the dual CFT at planar (leading large $N$) order, and its backreaction on the black hole \cite{Tanaka:2002rb,Emparan:2002px}. In \cite{Hubeny:2009ru} they were reconsidered without the UV cutoff, in which case they are dual to the CFT in a fixed black hole background. After several years of controversy, the existence of such regular static black hole solutions seems to be settled after the numerical construction in \cite{Figueras:2011va}. Nevertheless, a simpler solution to the problem, such as afforded by the large $D$ expansion, may be desirable.

We seek solutions for surfaces $\Sigma_B$ embedded in \eqref{PAdS} in the form
\beq
r=\bar{r}(z)\,.
\eeq
Choosing a normal that near the boundary points outwards from $\Sigma_B$ in the direction of increasing $r$, a direct computation of the mean curvature to leading order in $1/n$ gives
\beq\label{meanK}
\frac1{n}K=\frac{z+\bar{r}\bar{r}'}{\ell\bar{r}\sqrt{1+(\bar{r}')^2}}\,.
\eeq
In this case eq.~\eqref{effeq} becomes
\beq\label{req2}
\frac{\ell}{n z}K=\frac{z+\bar{r}\bar{r}'}{z\bar{r}\sqrt{1+(\bar{r}')^2}}=2\tilde\kappa\,.
\eeq
%
Note that $\ell$ disappears from this equation, as it must, without having set it to $1$.

Eq.~\eqref{req2} is a non-linear equation that requires numerical integration, but it is convenient to first analyze its main properties.
The equation simplifies for $\tilde{\kappa}= 0$ to
\beq\label{rbsinft}
\bar{r}\,\bar r'+z=0\,,
\eeq
which is solved by a circular profile,
\beq\label{circ}
\bar r^2+z^2=\textrm{const.}
\eeq
The extrinsic curvature $K$ of this solution vanishes: the curvature of the sphere is exactly cancelled by the cosmological curvature. 
Other simple solutions are: 
\begin{itemize}
\item[-] the AdS black brane, with $z=1/(2\tilde{\kappa})$ and $\bar r'\to+\infty$; 
\item[-] the black string solution, with $\bar r=1/(2\tilde{\kappa})$.
\end{itemize}

Eq.~\eqref{req2} is invariant under $z\to -z$, and it requires that $\bar{r}'$ vanish at the boundary $z=0$. Therefore, the surface $\Sigma_B$ always meets the boundary orthogonally. 

When extended into the bulk, 
we want $\Sigma_B$ to cap off smoothly at a finite value $z=z_m$ where $\bar{r}=0$. The smoothness is guaranteed by eq.~\eqref{req2}: for small $\bar{r}$ and large $|\bar{r}'|$, it becomes
\beq\label{cap}
\bar{r}|\bar{r}'|\approx \frac{z}{1+2\tilde\kappa z}\,,
\eeq
which shows that if $\Sigma_B$ closes off away from the boundary, $z>0$, it does it smoothly.

The solutions of \eqref{req2} are completely determined once the boundary radius
\beq
r_b=\bar{r}(z=0)
\eeq
is specified. Therefore, the solutions are parametrized by the surface gravity, \ie\ by $\tilde{\kappa}$, and by the radius at the boundary $r_b$. These parameters can be varied independently, but only their product
\beq\label{lambkr}
\lambda=2\tilde\kappa r_b
\eeq
is invariant under the scaling symmetry of the background. Therefore we obtain a one-parameter family of inequivalent solutions labeled by $\lambda$.

Observe that, even if the near-horizon geometries for different values of $\lambda$ are all locally equivalent, they will match to different far-zone geometries. For instance, if we fix $r_b$, then as $\tilde\kappa$ varies we get boundary black hole geometries with the same horizon radius but with different radial dependence in $g_{tt}$. Ref.~\cite{Hubeny:2009ru} proposed the existence of a one-parameter family of black droplets of this kind. Our construction realizes it in a natural way. 

When $\tilde{\kappa}$ is finite and non-zero we can rescale $t,r,z$ such that we effectively set
\beq
2\tilde{\kappa}=1\,,
\eeq
in which case the solutions are parametrized by $\lambda=r_b$.  This choice is slightly convenient. With it, the two branches of solutions for $\bar{r}'$ in eq.~\eqref{req2} are 
\beq\label{reqz}
\bar r'=-\frac{z}{\bar r}\frac{1\pm \sqrt{z^2+\bar r^2(1-z^2)}}{1-z^2}\,,
\eeq
Consider first the $-$ sign solution. Substituting in \eqref{meanK} one finds that
\beq\label{Kplus}
\frac1{n}K_{(-)}=\frac{z}{\ell}\,.
\eeq
Given our choice of normal in \eqref{meanK}, since $K>0$ the criterion discussed at the end of sec.~\ref{sec:effLO} tells us that the black hole exterior is in the direction of increasing $r$. Thus, the $-$ sign in \eqref{reqz} gives black droplets.

On the other hand, for the $+$ sign in \eqref{reqz} we obtain\footnote{To obtain this we assume that $0\leq z<1$, which is appropriate since we want solutions connected to the boundary.}
\beq
\frac1{n}K_{(+)}=-\frac{z}{\ell}\,,
\eeq
so this gives solutions where the exterior is inverted relative to the previous case. Therefore these are not black droplets. It is less clear whether these solutions are relevant. We discuss them briefly in appendix~\ref{app:cavities}.

\begin{figure}[t]
\begin{center}
\begin{tabular}{cc}
\includegraphics[width=4cm]{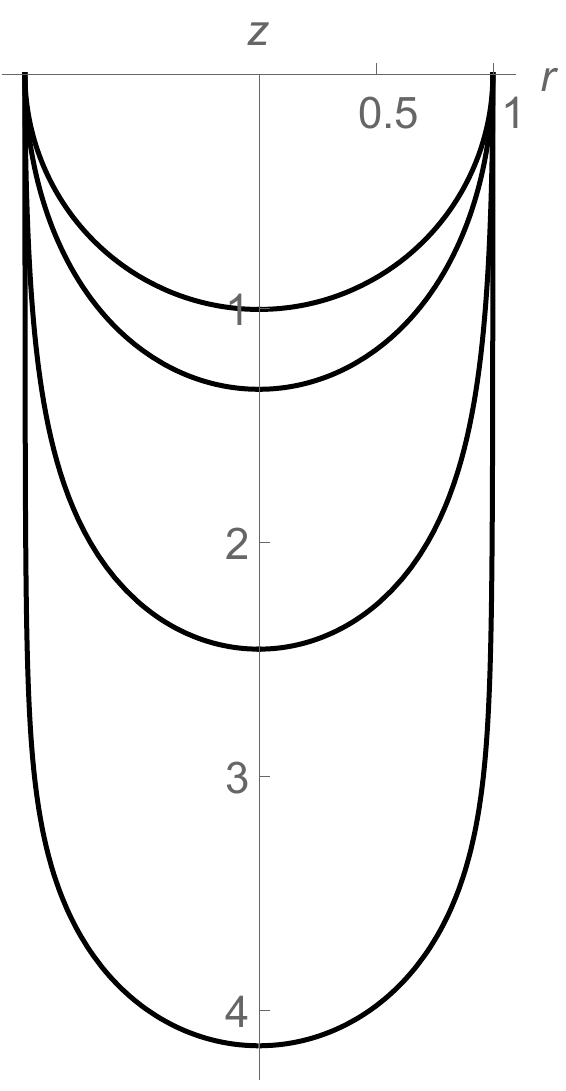}
\end{tabular}
\caption{\small Black droplet solutions for fixed horizon radius at the boundary ($r_b=1$) and varying surface gravity $\tilde\kappa$. The AdS boundary $z=0$ is at the top, and $z$ increases towards the bottom.  The solutions correspond to  $\lambda=2\tilde{\kappa} r_b=0.01,\, 0.5,\, 0.95,\, 0.9995$, with the elongation growing as $\lambda$ increases towards $1$.} 
\label{fig:droplet}
\end{center}
\end{figure}

Numerical integration of \eqref{reqz}, with the $-$ sign, yields the expected droplet-shaped solutions. In fig.~\ref{fig:droplet} we plot them with fixed $r_b=1$ and different values of $\tilde\kappa$. 

We see that as $\tilde\kappa$ grows with $r_b$ fixed, the droplets extend more into the bulk. The parameter $\lambda$ ranges between $0$ and $1$, and as $\lambda\to 0$ the solutions approach \eqref{circ}. Eq.~\eqref{rbsinft} is actually recovered from \eqref{req2} in the limit in which both $\bar r$ and $z$ go to zero at the same rate. At the opposite end, when $\lambda\to 1$, the solutions approach the black string. Near the black string limit, with $\lambda\lesssim 1$, the solution at small $z$ is close to a black string,
\beq
2\tilde\kappa\, \bar{r} =1-(1-\lambda) e^{2\tilde\kappa^2 z^2}\,.
\eeq
Close to the cap it is instead described by \eqref{cap}.

It is not fully clear from this study which of these black droplets should correspond to the solution whose boundary geometry is the Schwarzschild black hole, although a plausible candidate is the solution \eqref{circ}. This, as well as the calculation of the holographic stress-energy tensor of these solutions, requires a more detailed analysis which we postpone to the future.

\paragraph{Black droplets with black branes, and black funnels.}

In AdS one expects other classes of static black hole solutions that are related to droplets \cite{Hubeny:2009ru,Hubeny:2009kz}. Black droplets are often considered in conjunction with a black brane of infinite extent in directions parallel to the boundary. The two horizons are separated in the bulk, and in the large $D$ limit the interaction between them is suppressed exponentiallly in $D$. Then we can obtain these configurations by simply superimposing a flat black brane and one of our black droplets. Their interaction through graviton exchange across the bulk is non-perturbative in $1/D$ but in principle it is possible to incorporate such effects by computing the gravitational attraction between the two membranes with the effective stress-energy tensor of sec.~\ref{sec:effmem}.

`Black funnels' can be regarded as the merger of a black brane and a black string (or a long enough black droplet) that hangs from boundary. 
Our effective equations do not seem to allow to obtain these solutions. The apparent reason is that the `shoulder' at which the black string and the black brane are joined involves large gradients along the horizon, of order $D$, which fall outside the remit of the effective theory.

\section{Effective theory: next-to-leading order}\label{sec:NLOeq}

At the next order we must take into account the $\rho$-dependence in 
$\cK$ in \eqref{hatr}. It is convenient to separate it in the form
\beq\label{rsplit}
\cK(\rho,z)=\frac1{\sr_0(z)}\lp 1-\frac1{n}\delta\sr(\rho,z)\rp\,.
\eeq
Here $\sr_0(z)$ is different than the leading-order function $r_0(z)$ in \eqref{defr0}, since it also contains $1/n$ contributions. Since we will want to keep the horizon at $\rho=0$, we fix the ambiguity in this split by demanding that
\beq
\delta\sr(0,z)=0\,.
\eeq

We can obtain $\sr_0$ and $\delta\sr$ using eq.~\eqref{scacur} in \eqref{hatr} and taking into account that, at this order,
\beq
\frac1{N}\nabla^2 N=n\frac{\sD_a r_0 \sD^a\mc{R}}{r_0\,\mc{R}}\,.
\eeq
Then we get
\beqa
\frac1{\sr_0^2}&=&\frac{1}{r_0^2}-\frac{2p+3}{n}
\frac{1}{\mc{R}^2}\lp1-( \sD\mc{R})^2\rp-\frac{2}{n}\frac{\sD^2\mc{R}}{\mc{R}}-\frac{3}{n}\frac{\sD_a r_0 \sD^a\mc{R}}{r_0\,\mc{R}}+\frac{1}{n^2}{}^{(\gamma)}\!R\notag\\
&=&\frac{1}{r_0^2}+\frac{2p+3}{n}
\lp\frac1{\ell^2}-\frac1{r_0^2}\rp-\frac{2}{n}\frac{\sD^2\mc{R}}{\mc{R}}-\frac{3}{n}\frac{\sD_a r_0 \sD^a\mc{R}}{r_0\,\mc{R}}+\frac{1}{n^2}{}^{(\gamma)}\!R\,,\label{nlohr}
\eeqa
(recall that $n$ and $D$ are related by \eqref{ndp}) and
\beq
\delta\sr=\sr_1(z)\ln(\cosh(\rho/2))\,,
\eeq
with
\beq\label{defr1}
\sr_1(z)=\frac{2\sr_0^2}{\mc{R}^2}\lp 1-\sr_0^2 f^{ab}\sD_a\mc{R}\sD_b\mc{R}\rp\,,
\eeq
and  $f_{ab}$ given in \eqref{fab}. In these equations $\mc{R}(z)$ and the metric $\gamma_{ab}(z)$ with connection $\sD_a$ are the ones obtained in the previous section, which already had absorbed $1/n$ contributions from radial integrations. As with $\sr_0(z)$, they satisfy 
\beq
\sR(0,z)=\mc{R}(z)\,,\qquad 
\sg_{ab}(0,z)=\gamma_{ab}(z)\,. 
\eeq

If we rescale $\rho$ by a function $f(z)+g(\rho,z)/n$ and shift $z\to z+h(\rho,z)/n^2$, we can reach a gauge with
\beq
N(\rho,z)=\sr_0(z)\,.
\eeq
We now write
\beq\label{Knlo}
K=\frac{D-1}{\sr_0(z)}\coth\rho+\delta K(\rho,z)\,,
\eeq
where the $z$-dependence ambiguity is fixed by demanding that $\delta K(0,z)$ be finite. The radial equation for $\delta K$ is
\beq
\partial_\rho\delta K+2\coth\rho\, \delta K=-\frac{2\delta\sr}{\sr_0}\,,
\eeq
and its solution regular at $\rho=0$ is
\beqa
\delta K&=&-\frac2{\sr_0\sinh^2\rho}\int_0^\rho d\rho' \sinh^2(\rho') \delta\sr(\rho',z) \notag\\
&=&-\frac{\sr_1}{\sr_0}F_1(\rho)\,,\label{delK}
\eeqa
where we define
\beq
F_1(\rho)=\frac2{\sinh^2\rho}\int_0^\rho d\rho'\, \ln\lp\cosh(\rho'/2)\rp\sinh^2(\rho')\,.
\eeq

Next, setting
\beq
K^t{}_t=\frac{D-1}{\sr_0(z)\sinh\rho}+\delta K^t{}_t
\eeq
and using eq.~\eqref{Rtt} in \eqref{evolEq} we obtain the equation
\beq
\partial_\rho\delta K^t{}_t +\coth\rho\, \delta K^t{}_t=\frac{\sr_1}{\sr_0}\frac{F_1(\rho)}{\sinh\rho}+\sr_0\lp \frac1{r_0^2}-\frac{\sD_a r_0 \sD^a\cR}{r_0\,\mc{R}}\rp\,,
\eeq
which we integrate requiring regularity at $\rho=0$,
\beq
\delta K^t{}_t=\frac{\sr_1}{\sr_0}\frac{F_2(\rho)}{\sinh\rho}+\sr_0\lp \frac1{r_0^2}-\frac{\sD_a r_0 \sD^a\cR}{r_0\,\mc{R}}\rp\tanh(\rho/2)\,,
\eeq
with
\beq
F_2(\rho)=\int_0^\rho d\rho' F_1(\rho')\,.
\eeq

We can now integrate \eqref{extcurv} to obtain $g_{tt}=-\sV^2(\rho,z)$. If we write it in the form
\beq
\sV(\rho,z)=\sV_0(z)\tanh(\rho/2)\lp 1+\frac{1}{n}\delta V(\rho,z)\rp\,,
\eeq
with $\delta V(0,z)=0$, the result is
\beq
\delta V=\sr_1 F_3(\rho)+2\sr_0^2\lp \frac1{r_0^2}-\frac{\sD_a r_0 \sD^a\cR}{r_0\,\mc{R}}\rp\ln\cosh(\rho/2)\,,
\eeq
where
\beq
F_3(\rho)=\int_0^\rho d\rho'\frac{F_2(\rho')}{\sinh\rho'}\,.
\eeq

At this moment we can compute $\delta K^a{}_b$ and $\delta K^i{}_j$ and use them in the vector constraint to obtain the NLO corrections to the effective equation. We have done this, but a quicker route to the equation is to impose directly the condition that the surface gravity on the horizon is constant, which must hold at all orders. Both conditions can be seen to be equivalent.

Since we are fixing $\sV(0,z)=\sV_0(z)$ and $\sN(0,z)=\sr_0(z)$ 
at the horizon, the surface gravity is simply
\beq\label{surfnlo}
\kappa=\frac{D-1}{2}\frac{\sV_0(z)}{\sr_0(z)}\,.
\eeq
However, the effective theory uses functions of $z$ in the asymptotic overlap zone, $1\ll \rho\ll n$. In this region,
\beq
F_3(\rho)=\frac{1}{4}+\ord{e^{-\rho}}\,,
\eeq
so we find
\beq
\sV(\rho,z)=\sV_0(z)\lp 1+\frac{\sr_1(z)}{4n}\rp+\ord{\rho/n}+\ord{e^{-\rho}}\,.
\eeq
In order to match this to a value 
\beq
\sqrt{-g_{tt}}\bigl|_{\Sigma_B}=V(z)
\eeq  
computed in the far-zone background, we set\footnote{The subleading terms at $\ord{\rho/n}$ and $\ord{e^{-\rho}}$ have to match between zones if they solve the equations of motion, since we have fixed the radial gauge and these terms do not correspond to any collective degrees of freedom. See the related remarks in sec.~\ref{effstensor}.}
\beq\label{farV}
\sV_0(z)=V(z)\lp 1-\frac{\sr_1(z)}{4n}\rp\,.
\eeq

Finally, in the asymptotic zone we have $\sR(\rho,z)\bigl|_{\Sigma_B}=\mc{R}(z)$ and $\sg_{ab}(\rho,z)\bigl|_{\Sigma_B}=\gamma_{ab}(z)$. Then $\cK(\rho,z)\bigl|_{\Sigma_B}=1/\sr_0(z)$ and we can use its form \eqref{nlohr} in \eqref{surfnlo}.

\subsection{Effective equation at NLO}

Inserting \eqref{nlohr} and \eqref{farV} in \eqref{surfnlo} we obtain 
\beq\label{nloeq1}
\frac{4\kappa^2}{n(n+1)}=V^2(z)\lp\frac{1}{r_0^2}\lp 1-\frac{\sr_1}{2n}\rp +\frac{2p+3}{n\ell^2}-\frac{2}{n}\frac{\sD^2\mc{R}}{\mc{R}}-\frac{3}{n}\frac{\sD_a r_0 \sD^a\mc{R}}{r_0\,\mc{R}}+\frac{1}{n^2}{}^{(\gamma)}\!R\rp\,,
\eeq
where $r_0$ is the leading-order function given in \eqref{eqN}, namely,
\beq
r_0(z)=\frac{\cR(z)}{\sqrt{1-(\sD\cR(z))^2+\frac{\cR(z)^2}{\ell^2}}}\,.
\eeq
Using \eqref{ident} in \eqref{defr1}, after a short calculation this equation can finally be written as
\beq\label{nloeq2}
\frac{4\kappa^2}{n^2}=V^2(z)\lp \frac{1}{r_0^2}+\frac1{n}\lp \frac{2p+4}{\ell^2}-\frac{2\sD^2\cR}{\cR}-\frac{4\sD_a r_0\sD^a\cR}{r_0\,\cR}\rp+\frac1{n^2}{}^{(\gamma)}\!R^{ab}\lp\gamma_{ab}+\frac{\sD_a \cR \sD_b\cR}{\cR^2}\rp\rp\,,
\eeq
which is the effective equation at NLO we sought. The last terms, with the Ricci tensor of $\gamma_{ab}$, must be included in the second case of \eqref{assRab}.

In the particular case when $p=1$ this effective equation is
\beq\label{nloeq3}
\frac{4\kappa^2}{n^2}=V^2(z)\lp \frac{1}{r_0^2}+\frac1{n}\lp \frac{6}{\ell^2}-2\gamma^{zz}\frac{\cR''}{\cR}-(\gamma^{zz})'\frac{\cR'}{\cR}-4\gamma^{zz}\frac{r_0'\cR'}{r_0\,\cR}\rp\rp
\eeq
with
\beq
r_0(z)=\frac{\cR}{\sqrt{1-\gamma^{zz}(\cR')^2+\frac{\cR^2}{\ell^2}}}\,.
\eeq
These NLO equations admit as solutions the ones discussed in sec.~\ref{sec:ssoln}. Moreover, they also reproduce correctly the surface gravity of the corresponding exact black holes to this order.

It is easy to see that if $g_{tt}=1$ and the intrinsic curvature ${}^{(\gamma)}\!R^a{}_b$ vanishes (or can be neglected), then $\sr_1$ is a constant (for any $p$) when the leading-order equation is satisfied. In this case, which typically includes surfaces in the flat Minkowski background, the effective equation at NLO is equivalent to the soap-film equation
\beq\label{soapnlo}
K\bigl|_{\Sigma_B}=\mathrm{constant}
\eeq
expanded to NLO in $1/n$. In other cases, the covariant form of \eqref{nloeq2} is less straightforward.

\section{Non-uniform black strings}
\label{sec:nubs}

In the Minkowski background \eqref{minkback} the only solutions to 
the leading-order equation \eqref{effeqex} are the spherical Schwarzschild black hole and the uniform black strings. However, at finite values of $D$ it is known that non-uniform strings are possible \cite{Gubser:2001ac,Wiseman:2002zc,Kleihaus:2006ee}. Since they are absent from the leading-order large $D$ theory, we seek them using the equation \eqref{soapnlo} at the next-to-leading order. Moreover, the wavelength of Gregory-Laflamme perturbations in the large $D$ limit \cite{Asnin:2007rw,Emparan:2013moa} is $\sim \sqrt{n}$, which implies that non-uniform black strings must be sought among solutions of the second kind in \eqref{assgdR}.

Following these remarks, we write the background geometry as
\beq\label{nubsback}
ds^2=-dt^2+\frac{dz^2}{n}+dr^2+r^2 d\Omega_{n+1}\,,
\eeq
and take
\beq\label{nubsR}
\cR(z)=1+\frac{2\cP(z)}{n}\,.
\eeq
Here we have fixed the overall scale by setting the $S^{n+1}$-area-radius of the uniform string to one.

As a consequence of \eqref{nubsR}, we are limited to considering non-uniformities along the string of relative amplitude $\ord{1/n}$. Nevertheless, although small, these fluctuations are treated non-linearly.

For \eqref{nubsback} and \eqref{nubsR}, eq.~\eqref{nloeq3} takes the form
\beq
\cP''(z)+\cP(z)+\cP'(z)^{2}=\cP_{0}, \label{Meq2}
\eeq
where $\cP_0$ is related to the surface gravity $\kappa$ by
\beq
\kappa=\frac{n}{2}-\cP_0\,.
\eeq
%

Eq.~\eqref{Meq2} is non-linear in $\cP$, but already its linear approximation at small $\cP$ immediately reveals the existence of static, slightly non-uniform black strings with wavelength $\Delta z\simeq 2\pi$, which agrees with the value obtained in the small-amplitude perturbative analysis of the Gregory-Laflamme problem at $D\to\infty$ \cite{Asnin:2007rw,Emparan:2013moa}. 

Multiplying \eqref{Meq2} by $e^{2\cP(z)} \cP'(z)$, we obtain the first integral, 
\beq
\cP'(z)^{2}=-\cP(z)+\cP_0+\frac12 -k_1 e^{-2\cP(z)}\,. \label{RTeq}
\eeq 

It is useful to regard this equation as the classical mechanics of a particle in the potential
\beq
V(\cP)=\cP + k_1e^{-2\cP}\,,
\eeq
with the coordinate $z$ taking the role of time.
Both positive and negative values of $\cP$ are allowed as long as $|\cP|\ll n$. When $k_1>0$ the potential is bounded below and has a minimum at $\cP=\frac12\ln(2k_1)$. In this case there are solutions in which $\cP$ oscillates between two turning points: these solutions are non-uniform black strings. When $k_1\leq 0$ the potential is unbounded below for negative $\cP$, which is not acceptable since then $\cP$ will roll down to arbitrarily negative values, violating the condition $|\cP|\ll n$. Thus we only consider $k_1>0$.

We use the symmetry of eq.~\eqref{RTeq} under $\cP\to \cP+\alpha$, $\cP_0\to \cP_0+\alpha$, $k_1\to k_1 e^{2\alpha}$ to fix $k_1=1/2$
so the equation is
\beq\label{RTeq2}
\cP'(z)^{2}=\cP_0 -\cP(z)+\frac12 \lp 1-e^{-2\cP(z)}\rp\,.
\eeq
The constant $\cP_0$ corresponds to the energy of the particle and thus controls the range of oscillation of $\cP(z)$ \ie\ the amount of deformation. For $\cP_0=0$ we find the uniform black string with $\cP(z)=0$, and if $\cP_0\gtrsim 0$ we recover the small oscillations mentioned above. 

We are now interested in larger deformations, $1\ll \cP_0\ll n$. The potential in \eqref{RTeq2} has two regions. To the right, where $\cP>0$, it is dominated by the linear term $\sim \cP$. To the left, with $\cP<0$, it is dominated by the exponential $\sim e^{2|\cP|}$. In terms of particle motion, the particle will spend much more time (\ie\ long extent in $z$) moving relatively slowly in the right side of the potential, than in the left side where it will move very quickly (\ie\ very short extent in $z$) since the potential is very steep there. So the deformation of the string will have one large bulge (where $\cP>0$) extending over a long distance, and a smaller neck (where $\cP<0$) that extends over a small length.

We can make this quantitative by solving \eqref{RTeq2} in an approximate way for large $\cP_0$. In the right side of the potential, for $\cP>0$, we solve
\beq
\cP'(z)^{2}\approx \cP_0+\frac12 -\cP(z)
\eeq
\ie\ a parabolic profile
\beq\label{rightM}
\cP\approx \cP_0+\frac12-\frac14 (z-z_{max})^2\,,\qquad z_{max}=2\sqrt{\cP_0+\frac12}\,.
\eeq
Here $z_{max}$ measures the distance from the point $z=0$ of no-deformation, $\cP=0$, to the maximum deformation where $\cP=\cP_{max}\approx \cP_0$. In the left side, where $\cP<0$, we solve
\beq
\cP'(z)^{2}\approx \cP_0+\frac12\lp 1- e^{-2\cP(z)}\rp\,,
\eeq
\ie\
\beq\label{leftM}
\cP(z)\approx -\frac12 \ln\left[ \lp 2 \cP_0+1\rp\lp 1-\tanh^2(\sqrt{\cP_0+\frac12}(z+z_{min}))\rp\right]
\eeq
where
\beq
z_{min}=\frac{1}{\sqrt{\cP_0+\frac12}}\text{arctanh}\,\sqrt{1-\frac1{2\cP_0+1}}\simeq \frac1{2\sqrt{\cP_0}}\ln(8\cP_0)\,,
\eeq
is the length of the negative deformation from $z=0$ to the minimum where $\cP=\cP_{min}\approx -\frac12 \ln(2\cP_0)$.

The complete solution obtained by taking \eqref{rightM} for $\cP>0$ and \eqref{leftM} for $\cP<0$~\footnote{The match at $z=0$ is not completely smooth since the slopes differ at each side, but the difference is small for large $\cP_0$.}
gives an excellent approximation, as compared to numerical integration, to the entire profile of the deformation at large $\cP_0$. It also remains fairly good even for relatively small $\cP_0$ (in particular if we keep the $1/2$ offset for $\cP_0$ in \eqref{rightM} and \eqref{leftM}). We present the two types of calculation in fig.~\ref{fig:nubs}.

\begin{figure}[t]
\begin{center}
\begin{tabular}{cc}
\includegraphics[width=10cm]{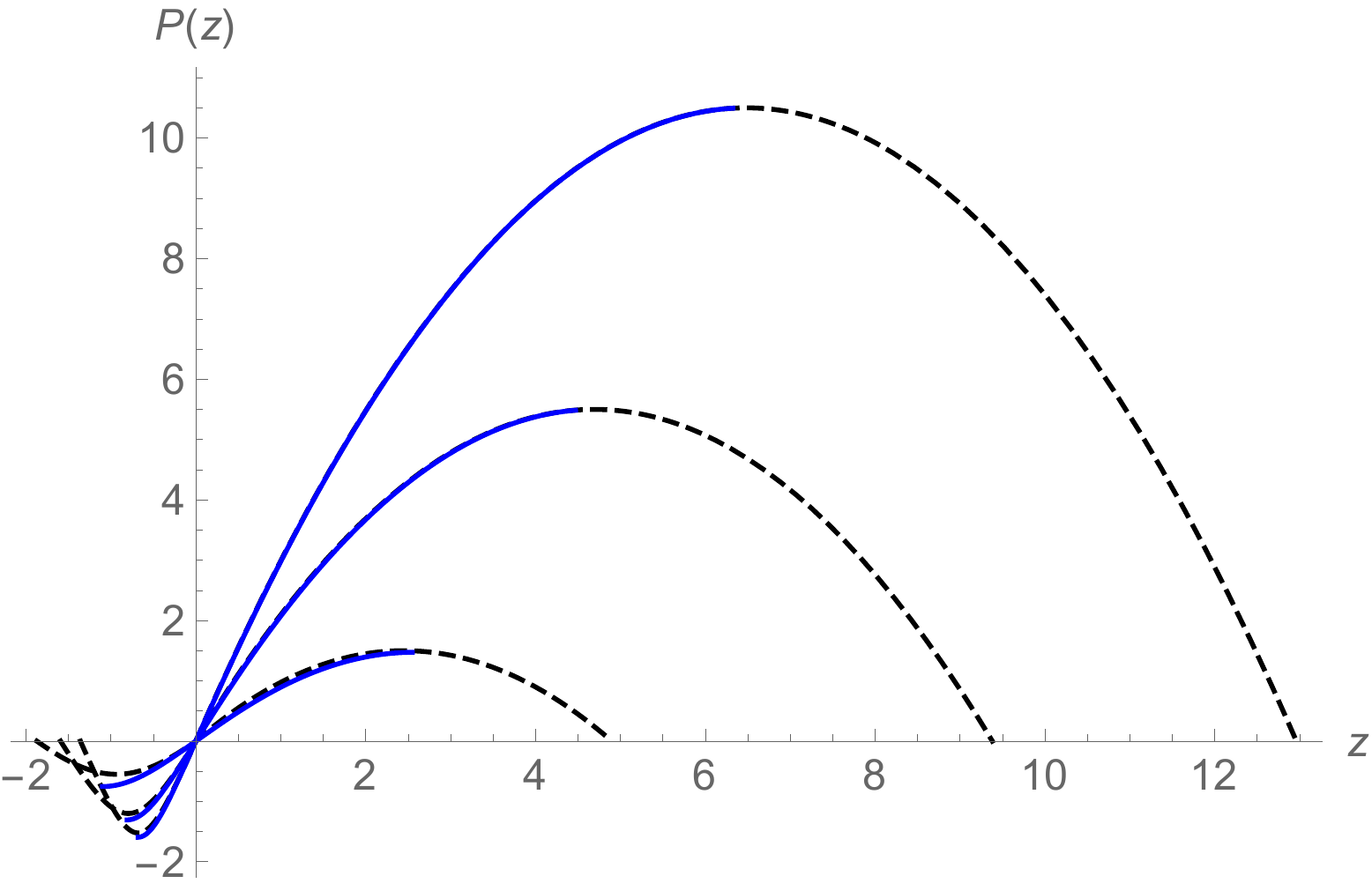}
\end{tabular}
\caption{\small Profile of the non-uniformity of black strings for increasing values of the deformation parameter $\cP_0=1,\,5,\,10$ (with $\cP_0=0$ corresponding to no deformation). Solid blue: numerical integration of \eqref{RTeq2}. Dashed black: analytic approximation \eqref{rightM} ($z>0$), \eqref{leftM}  ($z<0$). For clarity, we show only a half-period of the numerical results, and a full period of the analytical ones. The relative non-uniformity is $\Delta\cR/\cR=2\cP(z)/n$ and the proper length along the string is $z/\sqrt{n}$.}  
\label{fig:nubs}
\end{center}
\end{figure} 

Let us now investigate the breakdown of the large-$D$ expansion when the amplitude of the deformation grows large.
Using the proper coordinate 
\beq
Z=\frac{z}{\sqrt{n}}\,,
\eeq 
we find that, according to \eqref{rightM}, the large bulge has extent 
\beq
(\Delta Z)_\text{bulge}\approx 2\sqrt{\frac{\cP_0}{n}}\,,
\eeq 
whereas from \eqref{leftM},  the short neck has extent 
\beq
(\Delta Z)_\text{neck}\approx \frac1{2\sqrt{n\cP_0}}\ln(8\cP_0)\,.
\eeq
If we consider a deformation at the limit of validity of our approximations, $\cP_0\sim n$, then
\beq
(\Delta Z)_\text{bulge}= \Or(1)\,,\qquad (\Delta Z)_\text{neck}=\Or\lp \frac{\ln n}{n}\rp\,.
\eeq
Therefore, as the non-uniformity of the black string grows, it goes from having wavelength $\sim 1/\sqrt{n}$ and amplitude $\cP_0/n=\ord{1/n}$, to developing bulges of size $\cP_0/n=\ord{1}$, which resemble spherical black holes of $\ord{1}$ radius that are joined by thin necks. At these necks the approximation breaks down since the gradients are large, $\sim n/\ln n$.

\section{Outlook}\label{sec:outlook}

In this article we have shown that an effective theory of black holes using membrane-like variables can be constructed in the large $D$ limit. In spite of broad similarities to the membrane paradigm of \cite{membrane}, this theory differs from it in significant ways that may be worth spelling out. The basic idea in the membrane paradigm is to exploit large boosts near a null surface. Then it depends only the Rindler geometry. In contrast, the large-$D$ effective membrane does not reside at the black hole horizon but at the outer boundary of the near-horizon region, which contains much more structure of the geometry than the Rindler limit. The large $D$ theory is based on a neat separation between the black hole and the background, absent at finite $D$, which permits the explicit integration of short-wavelength degrees of freedom and gives an equation for the shape of the black hole. The membrane paradigm of \cite{membrane}, instead, does not yield any constraints nor information on the black hole shape. An equation for black hole horizons that is similar to \eqref{effeq0} has been derived, in any $D$, in \cite{Jaramillo:2013rda}, but it is unclear to us if there is a connection to our approach. 

As in the case of the effective black hole theories of \cite{Bhattacharyya:2008jc,Emparan:2009at}, finding solutions using the large-$D$ equations is considerably simpler than in the full Einstein theory. This is because: (i) the radial direction has already been integrated, resulting in a reduction by one in the cohomogeneity of the system, \ie\ in the effective dimensionality of the equations to solve; (ii) there is only one degree of freedom (for static black holes), hence one equation to solve, instead of a system of coupled differential equations. 

We have demontrated that two classes of non-trivial solutions are simply obtained with this method, but the detailed study of their physical properties has been outside the scope of this article. This includes the computation of the holographic stress-energy tensor for the black droplets, which requires to consider how the effective stress-energy tensor affects the far zone. The thermodynamic properties of the non-uniform black strings 
will be the subject of \cite{RSKTnubs}. 

Clearly, there is much scope for extending the theory. The generalization to stationary black holes, which have an additional collective coordinate corresponding to a local boost along the horizon, will be presented elsewhere \cite{RSKTrot}. Including charge for the black holes is similarly straightforward \cite{ADRE}. The natural next step is the more general theory that includes not only spatial fluctuations but also time evolution on timescales $\Delta t\gg r_0/D$  --- we understand that this is being pursued in \cite{shiraz}.

\section*{Acknowledgements}

This work was begun during the workshop ``Holographic vistas on Gravity and Strings'' YITP-T-14-1 at the Yukawa Institute for Theoretical Physics, Kyoto University, whose kind hospitality and support we all acknowledge. RE thanks the Galileo Galilei Institute for Theoretical Physics for the hospitality and the INFN for partial support during the completion of this work. 
RE is supported by FPA2010-20807-C02-02, FPA2013-46570-C2-2-P, AGAUR 2009-SGR-168 and CPAN CSD2007-00042 Consolider-Ingenio 2010. TS is supported by Grant-Aid for Scientific Research from Ministry of 
Education, Science, Sports and Culture of Japan (No.25610055). KT was supported by JSPS Grant-in-Aid for Scientific Research No.26-3387. TT was supported by MEXT Grant-in-Aid for Scientific Research on Innovative Areas, ``New Developments in  Astrophysics Through Multi-Messenger Observations of Gravitational Wave Sources'', Nos.\ 24103001 and 24103006.
Grant-in-Aid for Scientific Research (B) No.\ 26287044.


\addcontentsline{toc}{section}{Appendices}
\appendix

\section{Curvature on sections at constant $\rho$}\label{app:rhocurv}

For a metric of the form
\beq
g_{\mu\nu}dx^\mu dx^\nu=-\sV^2(\rho,z)dt^2+\sg_{ab}(\rho,z)dz^a dz^b+\sR^2(\rho,z) d\Omega_{n+1}\,,
\eeq
the Ricci tensor at any given value of $\rho$ is (exactly in $n$)
\beqa
R^t{}_t&=&-\frac{\bar{\sD}^2\sV}{\sV}-(n+1)\sg^{ab}\frac{\partial_a\sR\,\partial_b\sV}{\sR\,\sV}\,,\label{Rtt}\\
R^a{}_b&=&{}^{\sg}\!{R}^a{}_b-\frac{\bar{\sD}_a\bar{\sD}^b\sV}{\sV}-(n+1)\frac{\bar{\sD}_a\bar{\sD}^b\sR}{\sR}\,,\label{Rab}\\
R^i{}_j&=&\delta^i{}_j\lp \frac{n}{\sR^2}\lp 1-(\bar{\sD}\sR)^2\rp-\frac{\bar{\sD}^2\sR}{\sR}-\sg^{ab}\frac{\partial_a\sR\,\partial_b\sV}{\sR\,\sV}\rp\,,\label{Rij}
\eeqa
and the scalar curvature
\beq\label{scacur}
R=\frac{n(n+1)}{\sR^2}\lp 1-(\bar{\sD}\sR)^2\rp-2(n+1)\lp\frac{\bar{\sD}^2\sR}{\sR} +\sg^{ab}\frac{\partial_a\sR\,\partial_b\sV}{\sR\,\sV}\rp-2\frac{\bar{\sD}^2\sV}{\sV}+{}^{(\sg)}\!{R}\,.
\eeq
Here $\bar{\sD}$ and ${}^{(\sg)}\!{R}^a{}_b$ are the connection and Ricci tensor for the metric $\sg_{ab}$, and they are $\rho$-dependent. With our choice of fixing the $z$ dependence at the horizon, up to NLO they coincide with the $\sD$ and ${}^{(\gamma)}\!{R}^a{}_b$ of the metric $\gamma_{ab}$ both at the horizon and at the surface $\Sigma_B$ in the asymptotic overlap zone.

\section{Another derivation of the effective equation}\label{app:another}

In this appendix we take a quicker route to deriving (admittedly, with some hindsight) the effective equation \eqref{effeqex1}. Instead of solving the large $D$ Einstein equations \textit{ab initio}, as we have done in sec.~\ref{sec:LOtheory}, here we take an already known solution, from the large $D$ limit of Schwarzschild(-AdS), and make its parameters become slowly-varying functions along directions $z^a$ parallel to the horizon. Then we impose Einstein's equations to find the equations that the deformation functions must satisfy, and solve for any $\rho$-dependent correction to the metric required to obtain a solution.
Although this approach is less systematic and complete than the one followed in sec.~\ref{sec:LOtheory}, it affords a quicker route to the derivation of the equations that can be useful in other instances.

Consider the large $D$ limit of a Schwarzschild black $p$-brane
\beq
ds^2=-4\tilde\kappa^2 r_0^2\tanh^2(\rho/2) dt^2+\frac{r_0^2}{n^2}d\rho^2 +\delta_{ab}dz^a dz^b+\mc{R}^2_0\;(\cosh(\rho/2))^\frac{4}{n+1}d\Omega_{n+1}\,. 
\eeq
The solution is naturally parametrized by the sphere radius $\cR_0$, by the surface gravity, which we parametrize with the radius $r_0$, and by the metric along the $z$ directions. In order for this to be a solution, it must be that $r_0^{-2}-\mc{R}_0^{-2}=\ell^{-2}$ and that the metric along the $z$ directions be flat.

Now we promote these parameters to functions of $z$, taking the metric ansatz 
\beqa\label{deformbh}
ds^2&=&-V_0^2(z)\tanh^2(\rho/2) dt^2+\frac{r_0^2(z)}{n^2}d\rho^2+\lp\gamma_{ab}(z)+ \frac1{n}\delta\gamma_{ab}(\rho,z)\rp dz^a dz^b\notag\\
&&+\mc{R}^2(z)(\cosh(\rho/2))^\frac{4}{n}d\Omega_{n+1}\,,
\eeqa
where we make the same assumptions about the dependence on $\rho$ and $z$ as in sec.~\ref{sec:LOtheory}. We have added a term $\delta\gamma_{ab}(\rho,z)$ expecting that it will be required in order to obtain a solution.\footnote{Other terms could be added, but when solving the equations one easily sees that either they do not enter at leading order, or they must vanish.}

At this point we may proceed to impose the Einstein equations. However, we can anticipate that these equations will require uniform surface gravity on the horizon (for the Killing vector $\partial_t$). So, in the interest of expediency, we fix this already by imposing \eqref{effeq1}.

For obtaining the Einstein equations we perform a Kaluza-Klein reduction on the large-dimension sphere $S^{n+1}$. Consider a metric of the warped-product form
\beq
d\hat s^2= g_{\mu\nu}(x)dx^\mu dx^\nu +e^{2\phi(x)}d\Omega_{n+1}
\eeq
with $\mu,\nu=0,\dots,p+1$ so the total dimension is $p+n+3$. Hatted quantities are for the complete metric, unhatted ones for the $(p+2)$-metric. 
The Einstein-AdS action is
\beqa
I&=&\int d^{p+n+3}x \sqrt{-\hat g}\lp \hat R+\frac{(p+n+2)(p+n+1)}{\ell^2}\rp\notag\\
&=&\Omega_{n+1}\int d^{p+2}x \sqrt{- g}\,e^{(n+1)\phi}\biggl( R+n(n+1)\lp \lp\partial\phi\rp^2+e^{-2\phi}\rp\notag\\
&&\qquad\qquad\qquad\qquad\qquad\qquad\quad +\frac{(p+n+2)(p+n+1)}{\ell^2}\biggr)\,.
\eeqa

The Einstein equations along the directions $x^\mu$ are (see app.~\ref{app:rhocurv})
\beqa\label{eeom}
\hat G_\mu{}^{\nu}&=&G_\mu{}^{\nu}-(n+1)(\partial_\mu\phi \partial^\nu\phi +\nabla_\mu\partial^\nu\phi)\notag\\
&&+\frac12 \delta_\mu{}^{\nu}\lp (n+1)(n+2)\lp\partial\phi\rp^2+2(n+1)\Box\phi-n(n+1)e^{-2\phi}\rp\notag\\
&=&\frac{(p+n+2)(p+n+1)}{2\ell^2}\delta_\mu{}^{\nu}\,.
\eeqa
The remaining equation is the one for the dilaton
\beq\label{dileom}
\Box\phi +(n+1)\lp\partial\phi\rp^2-n e^{-2\phi}=\frac{p+n+2}{\ell^2}\,.
\eeq

Now we apply this to the metric \eqref{deformbh} and take the large $n$ limit. To illustrate the method more clearly, we restrict ourselves to the case $p=1$. 

From \eqref{deformbh} we have
\beq
\phi=\ln \mc{R}(z)+\frac{2}{n}\ln(\cosh(\rho/2))\,.
\eeq
It is straightforward to see how the different terms in the equations scale with $n$. To leading order one easily computes
\beq
\Box\phi= \frac{n}{r_0^2(y)\cosh^2(\rho/2)}\,,\qquad 
\lp\partial\phi\rp^2=\frac{\tanh^2(\rho/2)}{r_0^2(y)}
+(\sD\ln\cR)^2\,,
\eeq
so that \eqref{dileom} requires that
\beq\label{tteq}
\frac1{r_0(z)^2}-\frac1{\mc{R}(z)^2}+(\sD\ln\cR)^2=\frac{1}{\ell^2}\,,
\eeq
\ie\ the same as \eqref{eqN}, which is indeed the effective equation that one solves in order to determine the embedding of the surface
\beq
ds^2\bigl|_{\Sigma_B}=-4\tilde\kappa^2 r_0^2(z)dt^2+\gamma_{zz}(z)\,dz^2+\mc{R}^2(z)d\Omega_{n+1},
\eeq
in the background spacetime.

When this equation is satisfied, one easily sees that all the other equations \eqref{eeom} are also solved to leading order, except for the one for $\hat G_a{}^{\rho}=0$, since both $\partial_a\phi \partial^\rho\phi$ and $\nabla_a\partial^\rho\phi$ are $\Or(n)$ (unlike all other off-diagonal entries) and thus enter at leading order. This equation requires that
\beq
\partial_z\ln r_0=\partial_z \ln\mc{R} \lp 1-\frac{\partial_\rho \delta\gamma_{zz}}{2\tanh(\rho/2)}\rp\,.
\eeq
When $r_0$ and $\mc{R}$ are not constant, this is solved by
\beqa
\delta\gamma_{zz}&=&4 \lp 1-\frac{\partial_z\ln r_0}{\partial_z \ln\mc{R}}\rp\ln \cosh(\rho/2)\notag\\
&=&4 r_0^2\lp \frac{1}{\ell^2}-\frac{\sD^2\mc{R}}{\mc{R}}\rp\ln \cosh(\rho/2)\,,
\eeqa
where in the last line we have used \eqref{tteq}. We have also absorbed any integration function of $z$ into $\gamma_{zz}$ so that $\delta\gamma_{zz}(0,z)=0$.

In this form, we have arrived at the metric \eqref{effsoln} and the equation \eqref{effeqex1}.

\section{Far-zone Schwarzschild field from the effective stress tensor}\label{app:farSchw}

In this appendix we show how the effective stress tensor \eqref{intT} yields the correct field for the large-$n$ Schwarzschild solution, when this is obtained as a solution of the $p=1$ effective equation \eqref{minkback} of the form 
\beq
r_0(z)=r_h\,,\qquad
\mc{R}(z)=r_h\sqrt{1-\frac{z^2}{r_h^2}}\,.
\eeq
We want to recover the large $n$  linearized field of the Schwarzschild black hole as the field created by a line source (a string) along the segment $z\in [-r_h,r_h]$, with an energy density \eqref{intT} of the form
\beq\label{efflin}
\langle T_{tt}(z)\rangle=\frac{n\Omega_{n+1}}{16\pi G}r_h^{n} \lp 1-\frac{z^2}{r_h^2}\rp^{(n+1)/2}\,.
\eeq
At finite $n$ such a string segment would give rise not just to a spherically symmetric monopolar field but also to higher multipoles. However we will see presently that when $n\to\infty$ these multipoles become subdominant.

In the linearized approximation we need to consider the Newtonian potential created by this linear mass distribution. For a point mass source, the solution of
\beq
\nabla^2\Phi = 16\pi G M \delta^{(n+3)}(x)
\eeq 
in the cylindrical coordinates of \eqref{minkback} is
\beq
\Phi=-\frac{16\pi G M}{(n+1)\Omega_{n+2}}\frac1{(r^2+z^2)^{(n+1)/2}}\,.
\eeq
Hence, for the source \eqref{efflin} along the segment $y\in [-r_h,r_h]$
we have
\beq
\Phi(r,z)=-\frac{\Omega_{n+1}}{\Omega_{n+2}}r_h^{n} \int_{-r_h}^{r_h}d\hat z
\frac{(1-\hat z^2/r_h^2)^{(n+1)/2}}{\lp r^2+(z-\hat z)^2\rp^{(n+1)/2}}\,.
\eeq
We are in the far-zone outside $\Sigma_B$ and therefore at $r^2+z^2>r_h^2\lp 1+\ord{1/n}\rp$. Therefore in the region near the tip of the segment along $r=0$, we have $|z|-r_h>r_h/n$.
due to the numerator, at large $n$ the integrand is always strongly peaked around $\hat z^2/r_h^2\lesssim 1/n$ and we can use the saddle point approximation. A simple way to do the calculation is by writing
\beq
\frac{\hat z^2}{r_h^2} \approx \frac{x^2}{n}
\eeq
so that
\beqa\label{Phifin}
\Phi(r,y)&\approx& -\frac{\Omega_{n+1}}{\Omega_{n+2}}r_h^{n}\frac{r_h}{\sqrt{n}}\int_{-\infty}^{\infty}dx\frac{e^{-x^2/2}}{\lp r^2+(z-r_h x/\sqrt{n})^2\rp^{(n+1)/2}}\notag\\
&\approx& -\frac{\Omega_{n+1}}{\Omega_{n+2}}r_h^{n}\frac{r_h}{\sqrt{n}}\frac1{\lp r^2+z^2\rp^{(n+1)/2}}\int_{-\infty}^{\infty}dx\, e^{-x^2/2}\notag\\
&\approx &-\frac{r_h^{n+1}}{(r^2+z^2)^{(n+1)/2}}
\eeqa
where we have used that, at large $n$,
\beq
\Omega_{n+2}\approx \sqrt{\frac{2\pi}{n}}\,\Omega_{n+1}\,.
\eeq
Eq.~\eqref{Phifin} is the correct linearized field of the Schwarzschild black hole in cylindrical coordinates in $D=n+4$ dimensions. The essential point is that at large $n$ the main contribution comes from the energy density near $|z|\lesssim 1/\sqrt{n}$, \ie\ a point-like source, while the higher multipoles of field of the rest of the segment are strongly suppressed. 

\section{Black cavities?}\label{app:cavities}

Taking the $+$ sign in \eqref{reqz}, the exterior of the black hole is reversed relative to black droplets. We refer to these putative solutions as `black cavities', since they describe a spherically symmetric region of space enclosed inside a horizon. An example of what we mean by this is given by the deSitter foliation of AdS spacetime,\footnote{We thank Hong Liu for a discussion of this example.}
\beq\label{dSAdS}
ds^2=\frac{dr^2}{\frac{r^2}{\ell^2}+1}+\frac{r^2}{\ell^2}\lp -\lp 1-\frac{\rho^2}{\ell^2}\rp dt^2+\frac{d\rho^2}{1-\frac{\rho^2}{\ell^2}}+\rho^2 d\Omega_{n+1}\rp\,.
\eeq
This metric describes a region inside a horizon at $\rho=\ell$, which extends in the $r$-direction from the boundary at $r\to\infty$ until it closes off in the bulk at $r=0$. In \eqref{dSAdS}, however, there are no large radial gradients $\sim D$ close to the horizon, so this is generically different than the solutions we study here (even though $\rho=\ell$ is a minimal surface with $K=0$).

As in the case of black droplets, there is a one-parameter family of solutions to \eqref{reqz} with $+$ sign, labeled by $\lambda$ in \eqref{lambkr}. They range from $\lambda=0$ for the half-spherical cap \eqref{circ} to $\lambda\to\infty$ for the black brane at $z=1/(2\tilde\kappa)$. In fig.~\ref{fig:cavity} we plot some of them, keeping fixed the surface gravity and varying the boundary horizon radius $r_b$. As $r_b$ increases, the horizon grows in directions parallel to the boundary. The solutions resemble an AdS black brane with a horizon that is almost planar for $r\ll r_b$, but as $r$ approaches $r_b$, it bends towards the boundary and meets it at $r=r_b$. 

It is not clear whether such solutions can be completed into far-zone geometries that are regular, in particular along the line $r=0$ at the origin of the $S^{n+1}$. In principle one can fix  arbitrarily the geometry at the AdS boundary, and it may be possible to write down boundary geometries with the required behavior, namely, a boundary geometry that is regular on and inside a horizon of finite radius $r=r_h$, and which develops large radial gradients close to the horizon, as required for the large $D$ expansion. If such a boundary geometry can be found, it must admit an extension into the bulk. If this is regular in the cavity enclosed by the bulk horizon, then one might legitimately characterize it as a `black cavity'.

\begin{figure}[t]
\begin{center}
\begin{tabular}{cc}
\includegraphics[width=12cm]{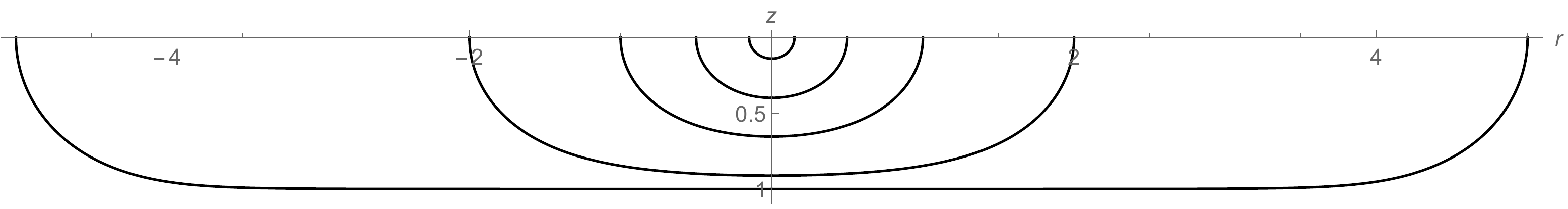}
\end{tabular}
\caption{\small `Black cavity' (?) solutions for fixed surface gravity ($\tilde{\kappa}=1/2$). The AdS boundary $z=0$ is at the top, and $z$ increases towards the bottom.  The exterior of the horizon is the region that extends towards \textit{smaller} $z$ (\ie\ reversed relative to fig.~\ref{fig:droplet}), so at the boundary, the exterior of the black hole is the region $0\leq r< r_b$. The solutions correspond to $r_b=0.15,\,0.5,\,1,\,2,\,5$.}  
\label{fig:cavity}
\end{center}
\end{figure}
%



\begin{thebibliography}{99}




\bibitem{Asnin:2007rw}
  V.~Asnin, D.~Gorbonos, S.~Hadar, B.~Kol, M.~Levi and U.~Miyamoto,
  ``High and Low Dimensions in The Black Hole Negative Mode,''
  Class.\ Quant.\ Grav.\  {\bf 24} (2007) 5527
  [arXiv:0706.1555 [hep-th]].

\bibitem{Emparan:2013moa}
  R.~Emparan, R.~Suzuki and K.~Tanabe,
  ``The large D limit of General Relativity,''
  JHEP {\bf 1306} (2013) 009
  [arXiv:1302.6382 [hep-th]].

\bibitem{Emparan:2013xia}
  R.~Emparan, D.~Grumiller and K.~Tanabe,
  ``Large D gravity and low D strings,''
  Phys.\ Rev.\ Lett.\  {\bf 110} (2013) 251102
  [arXiv:1303.1995 [hep-th]].

\bibitem{Emparan:2014cia}
  R.~Emparan and K.~Tanabe,
  ``Universal quasinormal modes of large D black holes,''
  Phys.\ Rev.\ D {\bf 89} (2014) 064028
  [arXiv:1401.1957 [hep-th]].

\bibitem{Emparan:2014jca}
  R.~Emparan, R.~Suzuki and K.~Tanabe,
  ``Instability of rotating black holes: large D analysis,''
  JHEP {\bf 1406} (2014) 106
  [arXiv:1402.6215 [hep-th]].

\bibitem{Emparan:2014aba}
  R.~Emparan, R.~Suzuki and K.~Tanabe,
  ``Decoupling and non-decoupling dynamics of large $D$ black holes,''
  JHEP {\bf 1407} (2014) 113
  [arXiv:1406.1258 [hep-th]].

\bibitem{Emparan:2015rva}
  R.~Emparan, R.~Suzuki and K.~Tanabe,
  ``Quasinormal modes of (Anti-)de Sitter black holes in the 1/D expansion,''
  arXiv:1502.02820 [hep-th].

\bibitem{Dias:2014eua}
  \'O.~J.~C.~Dias, G.~S.~Hartnett and J.~E.~Santos,
  ``Quasinormal modes of asymptotically flat rotating black holes,''
  Class.\ Quant.\ Grav.\  {\bf 31} (2014) 24,  245011
  [arXiv:1402.7047 [hep-th]].


\bibitem{Bhattacharyya:2008jc}
  S.~Bhattacharyya, V.~E.~Hubeny, S.~Minwalla and M.~Rangamani,
  ``Nonlinear Fluid Dynamics from Gravity,''
  JHEP {\bf 0802} (2008) 045
  [arXiv:0712.2456 [hep-th]].

\bibitem{Emparan:2009at}
  R.~Emparan, T.~Harmark, V.~Niarchos and N.~A.~Obers,
  ``Essentials of Blackfold Dynamics,''
  JHEP {\bf 1003} (2010) 063
  [arXiv:0910.1601 [hep-th]].

  J.~Camps and R.~Emparan,
  ``Derivation of the blackfold effective theory,''
  JHEP {\bf 1203} (2012) 038
   [JHEP {\bf 1206} (2012) 155]
  [arXiv:1201.3506 [hep-th]].

\bibitem{Gibbons:2002av}
  G.~W.~Gibbons, D.~Ida and T.~Shiromizu,
  ``Uniqueness and nonuniqueness of static black holes in higher dimensions,''
  Phys.\ Rev.\ Lett.\  {\bf 89} (2002) 041101
  [hep-th/0206049].

\bibitem{hsiang}
W.~Hsiang, Z.~Teng, W.~Yu,
``Examples of constant mean curvature immersions of the 3-sphere into euclidean 4-space,''
Proc.\ Natl.\ Acad.\ Sci.\ USA, {\bf 79} (1982) 3931.

\bibitem{Hubeny:2009ru}
  V.~E.~Hubeny, D.~Marolf and M.~Rangamani,
  ``Hawking radiation in large N strongly-coupled field theories,''
  Class.\ Quant.\ Grav.\  {\bf 27} (2010) 095015
  [arXiv:0908.2270 [hep-th]].

\bibitem{Gregory:1993vy}
  R.~Gregory and R.~Laflamme,
  ``Black strings and p-branes are unstable,''
  Phys.\ Rev.\ Lett.\  {\bf 70} (1993) 2837
  [hep-th/9301052].

\bibitem{Gubser:2001ac}
  S.~S.~Gubser,
  ``On nonuniform black branes,''
  Class.\ Quant.\ Grav.\  {\bf 19} (2002) 4825
  [hep-th/0110193].

\bibitem{Wiseman:2002zc}
  T.~Wiseman,
  ``Static axisymmetric vacuum solutions and nonuniform black strings,''
  Class.\ Quant.\ Grav.\  {\bf 20} (2003) 1137
  [hep-th/0209051].


\bibitem{Emparan:2011ve}
  R.~Emparan and N.~Haddad,
  ``Self-similar critical geometries at horizon intersections and mergers,''
  JHEP {\bf 1110} (2011) 064
  [arXiv:1109.1983 [hep-th]].
  
\bibitem{Tanaka:2002rb}
  T.~Tanaka,
  ``Classical black hole evaporation in Randall-Sundrum infinite brane world,''
  Prog.\ Theor.\ Phys.\ Suppl.\  {\bf 148} (2003) 307
  [gr-qc/0203082].

\bibitem{Emparan:2002px}
  R.~Emparan, A.~Fabbri and N.~Kaloper,
  ``Quantum black holes as holograms in AdS brane worlds,''
  JHEP {\bf 0208} (2002) 043
  [hep-th/0206155].

\bibitem{Figueras:2011va}
  P.~Figueras, J.~Lucietti and T.~Wiseman,
  ``Ricci solitons, Ricci flow, and strongly coupled CFT in the Schwarzschild Unruh or Boulware vacua,''
  Class.\ Quant.\ Grav.\  {\bf 28} (2011) 215018
  [arXiv:1104.4489 [hep-th]].


\bibitem{Hubeny:2009kz}
  V.~E.~Hubeny, D.~Marolf and M.~Rangamani,
  ``Black funnels and droplets from the AdS C-metrics,''
  Class.\ Quant.\ Grav.\  {\bf 27} (2010) 025001
  [arXiv:0909.0005 [hep-th]].

  V.~E.~Hubeny, D.~Marolf and M.~Rangamani,
  ``Hawking radiation from AdS black holes,''
  Class.\ Quant.\ Grav.\  {\bf 27} (2010) 095018
  [arXiv:0911.4144 [hep-th]].

  M.~M.~Caldarelli, \'O.~J.~C.~Dias, R.~Monteiro and J.~E.~Santos,
  ``Black funnels and droplets in thermal equilibrium,''
  JHEP {\bf 1105} (2011) 116
  [arXiv:1102.4337 [hep-th]].

  J.~E.~Santos and B.~Way,
  ``Black Funnels,''
  JHEP {\bf 1212} (2012) 060
  [arXiv:1208.6291 [hep-th]].

  S.~Fischetti, D.~Marolf and J.~E.~Santos,
  ``AdS flowing black funnels: Stationary AdS black holes with non-Killing horizons and heat transport in the dual CFT,''
  Class.\ Quant.\ Grav.\  {\bf 30} (2013) 075001
  [arXiv:1212.4820 [hep-th]].

  S.~Fischetti and J.~E.~Santos,
  ``Rotating Black Droplet,''
  JHEP {\bf 1307} (2013) 156
  [arXiv:1304.1156 [hep-th]].

  J.~E.~Santos and B.~Way,
  ``Black Droplets,''
  JHEP {\bf 1408} (2014) 072
  [arXiv:1405.2078 [hep-th]].


\bibitem{Kleihaus:2006ee}
  B.~Kleihaus, J.~Kunz and E.~Radu,
  ``New nonuniform black string solutions,''
  JHEP {\bf 0606} (2006) 016
  [hep-th/0603119].

  E.~Sorkin,
  ``Non-uniform black strings in various dimensions,''
  Phys.\ Rev.\ D {\bf 74} (2006) 104027
  [gr-qc/0608115].

  M.~Headrick, S.~Kitchen and T.~Wiseman,
  ``A New approach to static numerical relativity, and its application to Kaluza-Klein black holes,''
  Class.\ Quant.\ Grav.\  {\bf 27} (2010) 035002
  [arXiv:0905.1822 [gr-qc]].

  P.~Figueras, K.~Murata and H.~S.~Reall,
  ``Stable non-uniform black strings below the critical dimension,''
  JHEP {\bf 1211} (2012) 071
  [arXiv:1209.1981 [gr-qc]].

\bibitem{Jaramillo:2013rda}
  J.~L.~Jaramillo,
  ``A Young-Laplace law for black hole horizons,''
  Phys.\ Rev.\ D {\bf 89} (2014) 2,  021502
  [arXiv:1309.6593 [gr-qc]].

\bibitem{membrane} T.~Damour, ``Surface Effects in Black Hole Physics", Proceedings of the Second Marcel Grossmann Meeting on General Relativity, (edited by R. Ruffni, North Holland, 1982) p. 587.

  R.~H.~Price and K.~S.~Thorne,
  ``Membrane Viewpoint on Black Holes: Properties and Evolution of the Stretched Horizon,''
  Phys.\ Rev.\ D {\bf 33} (1986) 915.

\bibitem{RSKTnubs}
R.~Suzuki and K.~Tanabe, ``Non-uniform black strings and the critical dimension in the $1/D$ expansion", to appear.

\bibitem{RSKTrot}
R.~Suzuki and K.~Tanabe, ``Stationary black holes: Large D analysis", to appear.

\bibitem{ADRE}
A.~Di Dato and R.~Emparan, in progress.

\bibitem{shiraz}
S.~Bhattacharyya, A.~De, S.~Minwalla, R.~Prasad and A.~Saha, 
``A membrane paradigm at large $D$'', to appear.



\end{thebibliography}
\end{document}